\documentclass[pre,aps,twocolumn,superscriptaddress]{revtex4-2}

\usepackage{graphicx}
\usepackage{xcolor}
\usepackage{amssymb,amsfonts,amsmath}
\usepackage[hidelinks]{hyperref}

\newcommand{\kB}{k_{\rm B}}
\renewcommand{\vec}{\mathbf}
\newcommand{\rv}{\vec{r}}
\newcommand{\rmexc}{{\rm exc}}
\newcommand{\rmext}{{\rm ext}}

\newcommand{\sph}[1]{#1}
\newcommand{\sm}[1]{\bar{#1}}
\DeclareMathOperator{\sgn}{sgn}

\graphicspath{{./figures/}}

\begin{document}

\author{Stefanie M. Kampa}
\affiliation{Theoretische Physik II, Physikalisches Institut, Universit{\"a}t Bayreuth, D-95447 Bayreuth, Germany}
\author{Matthias Schmidt}
\email{Matthias.Schmidt@uni-bayreuth.de}
\affiliation{Theoretische Physik II, Physikalisches Institut, Universit{\"a}t Bayreuth, D-95447 Bayreuth, Germany}
\author{Florian Samm\"uller}
\affiliation{Theoretische Physik II, Physikalisches Institut, Universit{\"a}t Bayreuth, D-95447 Bayreuth, Germany}

\title{Spherical metadensity functional learning for inhomogeneous classical fluids}

\begin{abstract}
  We develop classical density functional learning to address fluids with truncated pairwise interparticle interactions in three-dimensional spherical geometry.
  Simulation data for systems with randomized repulsive pair potentials provide the basis for supervised training of a neural metadensity functional, thereby making efficient use of results for radial distribution functions in the bulk fluid via the test particle route.
  Specifically, we develop spherical local learning in order to represent the one-body direct correlation functional in terms of a neural network, which captures spatial curvature effects as well as the metadensity functional dependence on the thermally scaled pair potential.
  The framework yields efficient access to inhomogeneous structuring and related physical phenomena that occur in fluids and general solvents when adsorbed against curved solutes and confined inside of spherical and planar cavities.
  Test particle setups facilitate accurate prediction of the bulk fluid pair structure and verification of thermodynamic test particle sum rules via functional line integration.
  Applying the metadensity functional for Henderson inversion allows one to infer accurately the pair potential from the bulk radial distribution function.
  We address implications of the geometrical setup for two-body quantities and obtain the two-body direct correlation functional from automatic differentiation.
  For the hard sphere fluid, we confirm metadensity functional predictions against results from a standard neural density functional with fixed pair potential as well as to an analytic functional as given by fundamental measure theory.
  Simulation results provide further reference and corroborate reliable results of the spherical neural metadensity functional across a broad range of applications.
\end{abstract}

\date{June 15, 2026; revision: August 20, 2026}

\maketitle

\section{Introduction}
\label{SECintroduction}

Incorporating symmetry in an effective way is usually an important, if not indispensable, step towards the solution of any physical problem.
In soft matter, symmetry naturally emerges from the microscopic properties of the underlying classical many-body system.
In the unordered fluid phase, the arising equilibrium averages inherit their symmetry properties from the nature of the interparticle interactions as well as from the external environment under consideration.
Specifically, in an inhomogeneous setup, position-resolved thermal averages in a simple fluid will in general adhere to the geometry of an imposed external potential, provided that no symmetry breaking occurs, such as freezing into an ordered crystal.

Taking such geometrical features into account is paramount for making efficient and accurate predictions.
Instead of resolving the full three-dimensional geometry, a reduction to the \emph{relevant} spatial degrees of freedom is desired and comes in general with a significant reduction of conceptual and computational demands in any predictive scheme.
Leaving aside the homogeneous bulk fluid, planar symmetry arises as the arguably simplest setup, whereby all one-body quantites are resolved along a single spatial coordinate and remain translationally invariant in all perpendicular directions.
This type of symmetry is encountered naturally in soft matter when considering e.g.\ adsorption against planar substrates, confinement in channels with parallel walls, or interfaces separating coexisting fluid phases.

Spherical geometry, where the one-body structure of an inhomogeneous three-dimensional system is radially symmetric, constitutes another prototypical and highly relevant situation in simple and complex fluids.
Examples include the behavior around spherical solutes and inside of spherical pores, but also the bulk fluid itself when considering its correlation structure from Percus' test particle viewpoint \cite{Percus1962ApproximationMethodsClassical}, whereby one fluid particle is fixed at the origin and takes on the role of an external potential for all other particles.
When interparticle attraction is present, wetting and drying transitions of fluids are intricately affected by the presence of curved substrates \cite{Evans2003WettingCurvedSubstrates,Evans2004NonanalyticCurvatureContributions,Stewart2005WettingDryingCurved,Coe2022DensityDepletionEnhanced,Coe2023UnderstandingPhysicsHydrophobic}.
Being able to address spherical geometry is also relevant for the description of depletion forces between colloidal particles \cite{Gotzelmann1998DepletionForcesFluids,Roth1999DepletionForcesCurved}.
From a theoretical and computational standpoint, it is unsurprising that spherical geometry requires careful treatment due to the complications arising from curvature and the singular behavior encountered at the origin.

In machine learning, symmetry has also been exploited to a considerable extent.
Equivariant neural networks provide a systematic means of incorporating underlying geometrical properties of a considered problem directly in the neural network architecture \cite{Cohen2016GroupEquivariantConvolutional,Satorras2021EnEquivariantGraph}.
Besides constraining models to adhere to a specific symmetry group, learning symmetry from data constitutes a further promising approach for understanding and quantifying phenomena in physical applications \cite{Domina2026HowUnconstrainedMachinelearning}.
Identifying symmetry is hence a highly successful construction principle for powerful machine-learning-based methods.
As we summarize below, statistical mechanics in its classical density functional formulation is naturally suited for the incorporation of machine learning methods, thus giving rise to compelling schemes for predicting the behavior of fluids and soft matter.
Recent progress is also notable in the related fields of quantum density functional theory \cite{Pederson2022MachineLearningDensity,Huang2023CentralRoleDensity} and molecular simulation \cite{Unke2021MachineLearningForce}.

In a range of recent studies which we summarize below, the Mermin-Evans density functional map \cite{Mermin1965ThermalPropertiesInhomogeneous,Evans1979NatureLiquidvapourInterface} proved to be accessible via simulation-based machine learning.
Simulation data provides the ground truth for training of a \emph{neural functional}, which subsequentally facilitates accurate quantitative predictions.
As a surplus, the density functional underpinnings of the approach \cite{Evans1979NatureLiquidvapourInterface,Evans1992DensityFunctionalsTheory,Hansen2013TheorySimpleLiquids} give ready and unified access to a breadth of physical insight.
The variety of predicted physical phenomena thereby exceeds both quantitatively and qualitatively by far the limited scope of situations encountered during training.

While there are multiple ways to incorporate machine learning into classical density functional theory, the local learning strategy by Sammüller~{\it et~al.}~\cite{Sammuller2023NeuralFunctionalTheory,Sammuller2024WhyNeuralFunctionals,Sammuller2025NeuralDensityFunctional,Robitschko2025LearningBulkInterfacial,Sammuller2024HyperdensityFunctionalTheory,Sammuller2025WhyHyperdensityFunctionals,Sammuller2026DeterminingChemicalPotential} is a relatively simple yet powerful means to capture the nontrivial (over ideal gas) correlation effects.
The sole prerequisite of the method is being able to access simulation results for inhomogeneous density profiles $\rho(\rv)$, where $\rv$ denotes position, which then serve as the input for supervised training of a neural network that represents a suitably chosen density functional relationship.
Local density functional learning has proven to be extensible to a wide range of relevant systems and phenomena, including charged and dipolar fluids \cite{Bui2025FirstprinciplesApproachElectromechanics,Bui2025LearningClassicalDensity,Bui2026DielectrocapillarityExquisiteControl}, mixtures \cite{Zhou2026RolesBulkSurface,Robitschko2025LearningBulkInterfacial}, anisotropic and molecular fluids \cite{SimonMachineLearningDensity2024,CersonskyMachineLearningStatistical2025,Yang2025HighDimensionalOperator,Bui2026UnifiedMachineLearning}, as well as nonequilibrium settings \cite{Heras2023PerspectiveHowOvercome,Zimmermann2024NeuralForceFunctional}.

A considerable extension of the framework is metadensity functional theory \cite{Kampa2025MetadensityFunctionalTheory,Kampa2026MetadensityFunctionalLearning}, which gives rise to an extended machine learning scheme that yields access to the functional dependence on the interparticle interaction potential.
That the excess free energy functional depends on the interparticle interaction potential is a well-founded theoretical result with rigorous microscopic origin in the many-body statistical mechanics \cite{Evans1979NatureLiquidvapourInterface,Evans1992DensityFunctionalsTheory,Hansen2013TheorySimpleLiquids}.
For systems with pairwise interparticle interactions, the pertinent dependence is on the scaled pair potential $\beta\phi(r)$, where $r$ denotes interparticle distance and $\beta = 1 / (\kB T)$ with Boltzmann constant $\kB$ and absolute temperature $T$.
Getting to grips with this functional dependence has important consequences, arguably akin to those of capturing the functional dependence on the one-body density profile $\rho(\rv)$.
However, from the viewpoint of analytical density functional approaches, the metadensity dependence has remained merely formal, as making suitable approximations on paper proves to be difficult already for fixed form of $\phi(r)$.
A simplistic approach is given by the mean-field (random-phase) approximation, where the excess free energy functional is bilinear in density and linear in $\phi(r)$.
In common applications, the mean-field approximation is usually only employed to treat longer-ranged attraction or repulsion, with hard core repulsion being described by fundamental measure theory \cite{Rosenfeld1989FreeEnergyModel,Roth2010FundamentalMeasureTheory}; see also Refs.~\cite{Schmidt1999DensityfunctionalTheorySoft,Schmidt2000DensityFunctionalAdditive,Schmidt2000FluidStructureDensityfunctional,Schmidt2011IsometricMetamorphicOperations} which generalize fundamental-measure concepts to soft core behavior and recent work \cite{Belloni2026ClassicalDensityFunctional} to push the weighted-density approximation \cite{Denton1989ModifiedWeightedDensityFunctional} further.

Having tractable access to a universal metadensity functional hence bears important consequences, as demonstrated via the neural metadensity functional theory presented in Refs.~\cite{Kampa2025MetadensityFunctionalTheory,Kampa2026MetadensityFunctionalLearning} for the prototypical case of one-dimensional fluids.
Beyond direct predictive schemes, where one prescribes a desired form of $\phi(r)$, the genuine functional dependence on $\phi(r)$ can also be leveraged in order to address inverse design problems.
An iconic yet topical example is Henderson inversion \cite{Henderson1974UniquenessTheoremFluid}, where one seeks to identify the unique pair potential that corresponds to a prescribed form of the radial distribution function $g(r)$ at specified bulk density and temperature.
This inversion task from structural data to microscopic properties of a system, along with further generalizations, has been addressed successfully with neural metadensity functional theory for one-dimensional fluids \cite{Kampa2025MetadensityFunctionalTheory,Kampa2026MetadensityFunctionalLearning}.
In contrast to other common inversion methods, such as reverse Monte Carlo, the metadensity functional framework allows addressing Henderson inversion without engaging explicitly with the full many-body resolution.
Importantly, however, one needs to be able to resolve both $g(r)$ and $\phi(r)$ with respect to linear distance $r$ from the origin, which is trivial in one-dimensional geometry, but which necessitates further considerations in two and three dimensions.

Previous applications of classical density functional learning have usually been considered in reduced geometries, most commonly planar symmetry, in order to curb the computational demands caused by more general setups; we note that Glitsch {\it et al.}~\cite{Glitsch2025NeuralDensityFunctional} have demonstrated the feasibility of density functional learning in their two-dimensional hard disk system.
In the present work, we connect and augment several prior approaches in order to construct a neural density functional for three-dimensional fluids in spherical geometry.
We thereby make use of the metadensity functional concept and consider virtually arbitrary repulsive pair interaction potentials $\phi(r)$ of finite range, arguing that investigating this kind of functional dependence is particularly compelling in this specific geometric setup.
Our training scheme rests in part on data for radial distribution functions $g(r)$, which are accessible already in bulk fluid simulations and which are incorporated in a controlled way \cite{Dijkman2025LearningNeuralFreeEnergy,Ram2025LearnedFreeenergyFunctionals,Sammuller2024NeuralDensityFunctionals}.
Specifically, we aim to represent the radially resolved one-body direct correlation functional $c_1(r; [\rho,\beta\phi])$ via a neural network and capture the dependence on both the density profile $\rho(r)$ as a function of radial distance to the origin $r$ and the scaled pair potential $\beta \phi(r)$ as a function of interparticle distance~$r$; functional dependence is denoted by brackets.
The one-body direct correlation functional then provides exhaustive access to relevant thermodynamic and structural information, as we demonstrate.
In particular, the excess free energy functional $F_\rmexc[\rho,\beta\phi]$ is obtained by functional line integration of $c_1(r; [\rho,\beta\phi])$.
Automatic differentiation \cite{Baydin2018AutomaticDifferentiationMachine,Stierle2024ClassicalDensityFunctional} of $c_1(r; [\rho,\beta\phi])$ yields the two-body direct correlation functional $\sph{c}_2(r,r';[\rho,\beta\phi])$ in radial geometry.
We consider several consequences of the concrete availability of these fundamental objects for predicting physical phenomena and for addressing inverse design problems.

The paper is organized as follows.
We lay out the theoretical and methodological basis in Sec.~\ref{SECTheoryAndMethods} and give an overview of metadensity functional learning and neural functional concepts in Sec.~\ref{SECMetadensityFunctionalTheory}.
The adaptation of local learning to spherical geometry is described in Sec.~\ref{SECSphericalMetadensityFunctionalLearning}.
In Sec.~\ref{SECPairCorrelationFunctions}, we derive relations for pair correlation functions that are pertinent due to the spherical symmetry.
Our results are laid out in Sec.~\ref{SECResults}, starting in Sec.~\ref{SECTraining} with the training of a neural metadensity functional in spherical geometry.
We also show in Sec.~\ref{SECTraining} the feasibility of training a neural hard sphere functional based on spherically inhomogeneous data, which will serve as reference for the metadensity functional.
The neural functionals are applied for predictions of density profiles in Sec.~\ref{SECPrediction}, whereby we consider prototypical inhomogeneities in spherical and planar geometry.
We investigate in Sec.~\ref{SECHenderson} the utility of the neural metadensity functional for addressing Henderson inversion, i.e.\ for deducing the pair potential from a given form of the radial distribution function and thermodynamic state point.
In Sec.~\ref{SECPairCorrelationStructure}, the pair correlation structure in homogeneous and spherically heterogeneous environments is addressed using several different routes and functionals, including analytic fundamental measure theory.
In Sec.~\ref{SECTestParticleThermodynamics}, thermodynamic consistency of the spherial metadensity functional is verified via test particle sum rules.
We give our conclusions and an outlook in Sec.~\ref{SECConclusions}.

\section{Theory and methods}
\label{SECTheoryAndMethods}

\subsection{Metadensity functional theory and neural functionals}
\label{SECMetadensityFunctionalTheory}

\subsubsection{Density functional concepts}

We consider a range of equilibrium systems, indexed by $k$, in the grand ensemble with associated inverse temperature $\beta_k$, chemical potential $\mu_k$, external potential $V_{\rmext,k}(\rv)$, one-body density profile $\rho_k(\rv)$, and one-body direct correlation function $c_{1,k}(\rv)$.
The spatial position $\rv$ is kept general here, but we will specialize to spherical geometry in the subsequent sections.
We further consider pairwise interparticle interactions with a (repulsive) pair potential $\phi_k(r)$.
Density functional theory \cite{Evans1979NatureLiquidvapourInterface,Evans1992DensityFunctionalsTheory,Hansen2013TheorySimpleLiquids} ascertains that the one-body direct correlation functional $c_1(\rv;[\rho,\beta\phi])$ is unique, where the square brackets make explicit both the density functional dependence on $\rho(\rv)$ and the ``metadensity'' \cite{Kampa2025MetadensityFunctionalTheory} functional dependence on the thermally scaled pair potential $\beta\phi(r)$.
When considered for each system $k$ individually, we can express the corresponding one-body direct correlation function $c_{1,k}(\rv)$ as follows:
\begin{align}
  \label{EQc1generationGeneral}
  c_{1,k}(\rv) &= \ln\rho_k(\rv) + \beta_k \left(V_{\rmext,k}(\rv)-\mu_k\right),\\
  \label{EQc1trainingGeneral}
  c_1(\rv;[\rho_k,\beta_k\phi_k]) &= c_{1,k}(\rv).
\end{align}
Equation \eqref{EQc1generationGeneral} can be viewed as a general statistical mechanical identity \cite{Hansen2013TheorySimpleLiquids}, which does not make apparent the underlying functional dependence of $c_1(\rv)$ yet.
In contrast, Eq.~\eqref{EQc1trainingGeneral} is a key result of classical density functional theory \cite{Evans1979NatureLiquidvapourInterface,Evans1992DensityFunctionalsTheory,Hansen2013TheorySimpleLiquids} and states that the one-body direct correlation function of any system $k$ is already uniquely determined by its density profile $\rho_k(\rv)$ and its scaled pair interaction potential $\beta_k \phi_k(r)$.
Perhaps counter-intuitively, Eq.~\eqref{EQc1trainingGeneral} lacks any explicit dependence both on $V_{\rmext,k}(\rv)$ and on $\mu_k$, as one could naively expect on the basis of the right hand side of Eq.~\eqref{EQc1generationGeneral}.
The fact that $c_1(\rv;[\rho,\beta\phi])$ is a purely intrinsic object, which is universal across all possible inhomogeneities and thermodynamic state points, is crucial for the wide applicability of density funtional methods \cite{Evans1979NatureLiquidvapourInterface,Evans1992DensityFunctionalsTheory,Hansen2013TheorySimpleLiquids}.
In particular, acknowledging the density functional dependence of $c_1(\rv;[\rho,\beta\phi])$ turns Eq.~\eqref{EQc1generationGeneral} into an implicit (Euler-Lagrange) equation for the density profile, which is amenable to performant self-consistent solution schemes, provided that the functional form of $c_1(\rv;[\rho,\beta\phi])$ is accurately available.

The one-body direct correlation functional is generated by the intrinsic excess free energy functional $F_\rmexc[\rho,\beta\phi]$ according to
\begin{equation}
  \label{EQc1FromFexc}
  c_1(\rv;[\rho,\beta\phi]) = - \left. \frac{\delta \beta F_\rmexc[\rho,\beta\phi]}{\delta\rho(\rv)} \right|_{\beta\phi}.
\end{equation}
In the notation we have made explicit that the scaled pair potential $\beta\phi(r)$ is kept fixed in the functional differentiation with respect to $\rho(\rv)$.
The excess free energy functional is the nontrivial intrinsic contribution to the grand potential
\begin{equation}
  \label{EQOmega}
  \Omega[\rho, \beta\phi] = F_\mathrm{id}[\rho] + F_\rmexc[\rho, \beta\phi] + \int d\rv \rho(\rv) \left(V_\rmext(\rv) - \mu\right),
\end{equation}
with ideal free energy $F_\mathrm{id}[\rho] = \int d\rv \rho(\rv) \left( \ln\rho(\rv) - 1 \right)$; the thermal de Broglie wavelength has been set to unity.

Applying further functional density derivatives to Eq.~\eqref{EQc1FromFexc} yields a hierarchy of higher-order direct correlation functionals which give access to the many-body correlation structure of the fluid.
On the pair correlation level, the two-body direct correlation functional is defined as
\begin{equation}
  \label{EQc2Fromc1}
  c_2(\rv,\rv';[\rho,\beta\phi]) = \left. \frac{\delta c_1(\rv;[\rho,\beta\phi])}{\delta\rho(\rv')} \right|_{\beta\phi}
\end{equation}
and this forms a central object in the theory of inhomogeneous fluids.
In particular, the inhomogeneous two-body Ornstein-Zernike equation
\begin{equation}
  \label{EQOZinhom}
  h_2(\rv, \rv') = c_2(\rv, \rv') + \int d\rv'' h_2(\rv, \rv'') \rho(\rv'') c_2(\rv'', \rv')
\end{equation}
defines the relation of $c_2(\rv, \rv')$ to the inhomogeneous total pair correlation function $h_2(\rv, \rv')$ in the form of an integral equation.

\subsubsection{Local density functional learning}

The general workflow of neural density functional learning \cite{Sammuller2023NeuralFunctionalTheory,Sammuller2024WhyNeuralFunctionals} can be viewed as a two-stage process based on Eqs.~\eqref{EQc1generationGeneral} and \eqref{EQc1trainingGeneral}.
At the first stage, training data is gathered by simulation-based sampling of inhomogeneous equilibrium density profiles $\rho_k(\rv)$, e.g.\ using grand canonical Monte Carlo simulations.
This is performed for given values of $\beta_k$, $\mu_k$, and $V_{\rmext,k}(\rv)$, which are in practice all randomized for each individual system $k$.
Simple postprocessing of the thus obtained set of inhomogeneous density profiles makes all quantities on the right-hand side of Eq.~\eqref{EQc1generationGeneral} known and yields the one-body direct correlation function $c_{1,k}(\rv)$ for each system $k$ and for all positions $\rv$ where $\rho_k(\rv) > 0$.
This scheme can readily be extended to also probe the metadensity functional dependence by randomizing the form of $\phi_k(r)$ in each simulation, as demonstrated in Refs.~\cite{Kampa2025MetadensityFunctionalTheory,Kampa2026MetadensityFunctionalLearning} for \emph{one-dimensional} fluids.

The second stage consists of training a \emph{neural functional}, i.e.\ a neural network which represents $c_1(\rv;[\rho,\beta\phi])$, in order to achieve the matching condition \eqref{EQc1generationGeneral}.
In practice, one defines a loss function $\mathcal{L} = \sum_k \Vert c_{1,k}(\rv) - c_1(\rv;[\rho_k,\beta_k\phi_k]) \Vert_2$ that measures the discrepancy of the neural functional predictions $c_1(\rv;[\rho_k,\beta_k\phi_k])$ to the ground-truth simulation data $c_{1,k}(\rv)$ throughout all systems $k$ via the $L_2$ norm $\Vert \cdot \Vert_2$.
Machine learning libraries provide high-level tools to implement the loss function $\mathcal{L}$, the neural network, and the optimization of neural network parameters via backpropagation.
Note that the problem under consideration is a standard regression task.
Centering the machine learning workflow around the one-body direct correlation functional constitutes the arguably simplest computational approach, as $c_1(\rv)$ carries the role of both a universal density functional, cf.\ Eq.~\eqref{EQc1trainingGeneral}, and of a quantity that is immediately accessible in many-body simulations, cf.\ Eq.~\eqref{EQc1generationGeneral}.
We note that working instead with $F_\rmexc[\rho,\beta\phi]$ or $c_2(\rv,\rv';[\rho,\beta\phi])$, see Eqs.~\eqref{EQc1FromFexc} and \eqref{EQc2Fromc1}, is equally valid albeit slightly more involved due to the increased complexity of accessing these objects in simulations \cite{Dijkman2025LearningNeuralFreeEnergy,Ram2025LearnedFreeenergyFunctionals,Sammuller2024NeuralDensityFunctionals}.

The concrete strategy of representing $c_1(\rv;[\rho,\beta\phi])$ as a neural network still needs to be specified.
While there are different possibilities \cite{Sammuller2024NeuralDensityFunctionals,Dijkman2025LearningNeuralFreeEnergy,Ram2025LearnedFreeenergyFunctionals,Glitsch2025NeuralDensityFunctional}, a multilayer perceptron is the simplest neural network architecture and consists of an input layer, several fully connected hidden layers and an output layer.
We take this most general form as a universal approximator of the functional mapping that we seek to extract from the simulation data.
The nature of direct correlation functions gives further motivation for a specific type of input-output pairing:
Instead of mapping the entire density profile $\rho(\rv)$ to the entire one-body direct correlation function $c_1(\rv)$, we employ the concept of \emph{local learning} \cite{Sammuller2023NeuralFunctionalTheory,Sammuller2024WhyNeuralFunctionals,Sammuller2024NeuralDensityFunctionals,Sammuller2025NeuralDensityFunctional,Sammuller2026DeterminingChemicalPotential}, whereby, for short-ranged interparticle forces, the density profile is considered only in the vicinity of a specific position of interest, at which the neural network shall output the corresponding \emph{value} of the one-body direct correlation function.
The local learning approach relies on the fundamental insight \cite{Evans1979NatureLiquidvapourInterface,Hansen2013TheorySimpleLiquids} that direct correlations are generally short-ranged in fluids where particles interact only via short-ranged forces.
Exploiting this `nearsightedness principle' is hence crucial to achieve optimal data utilization and to conceptualize simple yet powerful neural network representation of the desired density functional relationship.
A substantial practical benefit of locally learned neural functionals is their multiscale applicability due to decoupling the neural functional input from the training system size \cite{Sammuller2023NeuralFunctionalTheory,Sammuller2025NeuralDensityFunctional,Bui2026UnifiedMachineLearning}.

In planar geometry, where inhomeneities are only encountered along a single axis denoted by $x$, local learning is both efficient and straightforward to implement, as the geometrical setup allows for a simple sectioning of the entire profiles $\rho_k(x)$ and $c_{1,k}(x)$ into corresponding pairs of density windows and one-body direct correlation values \cite{Sammuller2023NeuralFunctionalTheory,Sammuller2024WhyNeuralFunctionals}.
However, hypothetical brute force application of this scheme to three-dimensional problems would require full volumetric resolution of all systems under consideration.
In general, this can be prohibitively expensive from a computational standpoint both in terms of data generation as well as for neural network construction and utilization.
We note that optimizations of the neural network architecture may mitigate this limitation to a certain extent, as demonstrated in Ref.~\cite{Glitsch2025NeuralDensityFunctional} for a two-dimensional hard disk system.

As an entirely different approach, we consider in this work the generalization to spherical geometry, which allows one to work with a single spatial coordinate $r$ corresponding to the radial distance from the origin.
In contrast to planar symmetry, this spherical setup necessitates to take into account the curvature effects which are most pronounced near the origin where $r$ is small.
We emphasize that planar geometry is still contained as a limiting case when considering the behavior at large distances $r \rightarrow \infty$ from the origin, where curvature becomes negligible.
In the following, we lay out our machine learning strategy for dealing with spherical symmetry and for incorporating the metadensity dependence on the scaled pair potential $\beta\phi(r)$, which proves to be highly beneficial for subsequent applications.

\subsection{Spherical metadensity functional learning}
\label{SECSphericalMetadensityFunctionalLearning}

\subsubsection{Training data}

A significant part of our training data consists of \emph{test particle} \cite{Percus1962ApproximationMethodsClassical,Hansen2013TheorySimpleLiquids} density profiles $\rho_k(r)$ for corresponding randomized interparticle pair potentials $\phi_k(r)$ and randomized thermodynamic parameters $\beta_k$, $\mu_k$, where the index $k$ enumerates the simulated systems.
In accordance with the test particle concept of fixing one fluid particle at the origin, we set the external potential $V_{\rmext,k}(\rv)=\phi_k(r)$, rendering it effectively an external potential for all other particles.
Additionally, we recall that the pair distribution functions $g_k(r)$ in the \emph{bulk} fluid are connected to test particle density profiles $\rho_k(r) = \rho_{b,k} g_k(r)$ upon multiplication with the respective bulk density $\rho_{b,k}$.
Specializing Eqs.~\eqref{EQc1generationGeneral} and \eqref{EQc1trainingGeneral} to test particle geometry, thereby taking into account the above considerations, yields:
\begin{align}
  \label{EQsphericalTestParticleTrainingData}
  c_{1,k}(r) &= \ln(\rho_{b,k} g_k(r)) + \beta_k \left(\phi_k(r) - \mu_k\right),\\
  \label{EQsphericalTestParticleMatching}
  c_1(r;[\rho_k,\beta_k\phi_k]) &= c_{1,k}(r).
\end{align}
For the generation of a suitable training data set, we can refrain from simulating inhomogeneous systems in which $V_{\rmext,k}(r) = \phi_k(r)$ is imposed explicitly.
Instead, Eq.~\eqref{EQsphericalTestParticleTrainingData} only requires the quantities $g_k(r)$ and the bulk density $\rho_{b,k}$ as input, which are accessible already in bulk fluid simulations.
Analogously, a specialization of Eqs.~\eqref{EQc1generationGeneral} and \eqref{EQc1trainingGeneral} to the spatially constant bulk densities $\rho_k(r) = \rho_{b,k}$ of all systems results in
\begin{align}
  \label{EQsphericalBulkTrainingData}
  c^b_{1,k} &= \ln(\rho_{b,k}) - \beta_k \mu_k,\\
  \label{EQsphericalBulkMatching}
  c_1(r;[\rho_k,\beta_k\phi_k]) &= c^b_{1,k},
\end{align}
which will serve as an additional matching condition for the respective bulk values $c^b_{1,k}$ of the one-body direct correlation function in each system.

As a secondary training set, we use density profiles in planar geometry where values of $\beta_k$ and $\mu_k$ are again randomized together with randomized form of the planar external potential $V_{\rmext,k}(x)$; the label $k$ is re-used to index these additional training systems.
The associated one-body direct correlation function \eqref{EQc1generationGeneral} in planar geometry, together with the matching condition \eqref{EQc1trainingGeneral} are then given respectively by
\begin{align}
  \label{EQplanarTrainingData}
  c_{1,k}(x) &= \ln\rho_k(x) + \beta_k \left(V_{\rmext,k}(x) - \mu_k\right),\\
  \label{EQplanarMatching}
  c_1(r_\infty + x;[\rho_k;\beta_k\phi_k]) &= c_{1,k}(x).
\end{align}
In order to recover planar symmetry, we offset the positional argument of the neural functional by $r_\infty$, which is a radial reference position at (virtually) infinite separation distance from the origin, such that curvature effects of the spherical coordinate system vanish.
Further details on the practical implementation are laid out below.

Lastly, we also describe the strategy to add data for spherically inhomogeneous systems with randomized spherical potential $V_{\rmext,k}(r)$ and randomized interparticle potential $\phi_k(r)$.
Then for spherical distance $r$ the corresponding one-body direct correlation function \eqref{EQc1generationGeneral} and matching condition \eqref{EQc1trainingGeneral} are given respectively by
\begin{align}
  \label{EQinhomTrainingData}
  c_{1,k}(r) &= \ln\rho_k(r) + \beta_k \left(V_{\rmext,k}(r) - \mu_k\right),\\
  \label{EQinhomMatching}
  c_1(r;[\rho_k,\beta_k\phi_k]) &= c_{1,k}(r).
\end{align}
Note that these data cover a wider class than Eq.~\eqref{EQsphericalTestParticleTrainingData}, as in general $V_{\rmext,k}(r) \neq \phi_k(r)$ in Eq.~\eqref{EQinhomTrainingData}, which induces density profiles different from the test particle profiles $\rho_{b,k} g_k(r)$.
Despite the possibility of providing general inhomogeneous data, we refrain from doing so in the training of the neural metadensity functional, where we work solely with data in planar and test particle geometry, as obtained via Eqs.~\eqref{EQsphericalTestParticleTrainingData}, \eqref{EQsphericalBulkTrainingData}, and \eqref{EQplanarTrainingData}.
We will use inhomogeneous training data via Eqs.~\eqref{EQplanarTrainingData} and \eqref{EQinhomTrainingData} instead only for training of a neural functional for the hard sphere fluid without metadensity dependence as a means of comparison and cross-validation.

\subsubsection{Local learning in spherical geometry}

\begin{figure}[tb]
  \includegraphics[width=\columnwidth]{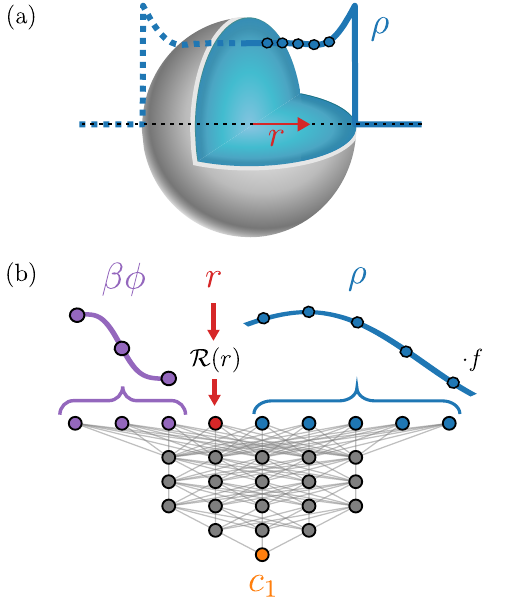}
  \caption{
    Local learning in spherical geometry.
    (a) Sketch of the geometrical setup for the exemplary case of a fluid in spherical confinement.
    The resulting density profile $\rho(r)$ is resolved (discretized) along the radial coordinate $r$ and symmetrically extended for negative values of $r$, such that $\rho(r) = \rho(-r)$.
    (b) The neural network consist of an input layer, several hidden layers (gray), and an output layer (orange), which yields the value of the one-body direct correlation functional $c_1(r; [\rho, \beta\phi])$ at the considered radial position.
    As input, we provide the discretized scaled pair potential $\beta \phi(r)$ and the discretized density window $\rho(r')$ for radial distances $r'$ within a predefined cutoff from the location of interest.
    The density window is thereby multiplied with the geometric weight function $f(r, r')$; see text.
    Additionally, we supply the scalar value $\mathcal{R}(r)$, i.e.\ the radial position $r$ at which a prediction is made, normalized by the function $\mathcal{R}$ prior to input.
  }
\label{FIGsphericallearning}
\end{figure}

Local learning for short-ranged interparticle interactions relies on finite input `windows' that represent the density functional dependence of $c_1(r;[\rho,\beta\phi])$.
In the present spherical geometry this implies to consider, for given distance $r$ from the origin, a density interval $[r-r_w,r+r_w]$, where $r_w$ is typically of the order of (several) particle sizes \cite{Sammuller2023NeuralFunctionalTheory,Sammuller2024WhyNeuralFunctionals,Sammuller2024NeuralDensityFunctionals,Sammuller2025NeuralDensityFunctional,Kampa2025MetadensityFunctionalTheory,Kampa2026MetadensityFunctionalLearning}.
Care is required to treat the vicinity of the origin, where $r$ is small such that $r - r_w$ may become negative.
We describe our strategy in the following.

Each radial training density profile $\rho(r)$ is mirrored at the origin, such that we \emph{define} values at negative separation $r < 0$ via $\rho(-r) = \rho(r)$.
This generates an even continuation of the density profile and implies continuity at $r = 0$.
The mirroring method allows one to work with input density windows of fixed size $2 r_w$, irrespective of the radial distance $r$ from the origin, in keeping with the general setup used in planar geometry \cite{Sammuller2023NeuralFunctionalTheory,Sammuller2024WhyNeuralFunctionals}.

As an additional means to establishing the mirror symmetry at the origin, we symmetrize the one-body direct correlation function as follows:
\begin{equation}
  \label{EQsphericalc1Symmetrized}
  c_1(r;[\rho,\beta\phi]) \leftarrow \frac{c_1(r;[\rho,\beta\phi]) + c_1(-r;[\rho,\beta\phi])}{2},
\end{equation}
where the arrow indicates that we replace the raw prediction of the neural functional by the average of the predictions for positive and negative separation distance.
Equation~\eqref{EQsphericalc1Symmetrized} constitutes a mere quantitative polishing step, as we find that the mirror symmetry is already acquired satisfactorily in the training.

Further crucial extensions of local learning concern the construction of the neural network and the preprocessing of input quantities.
We depict an illustration of the neural network and our geometrical setup in Fig.~\ref{FIGsphericallearning}.
Translational invariance along the radial $r$ axis is broken by construction, as curvature effects vary depending on the distance to the origin; we recall that $r \rightarrow \infty$ corresponds to vanishing curvature and hence recovers the case of planar geometry.
We hence extend the neural network input layer and encode the curvature information in an additional (scalar) input node.
The corresponding input quantity is $\mathcal{R}(r)$, i.e.\ the value of $r$ is transformed prior to input by a suitable normalizing function $\mathcal{R}$.
For the neural metadensity functional, we choose $\mathcal{R}(r) = a \sgn(r)((\tanh(|r|/2-1)+1)/2-(\tanh(-1)+1)/2)$ heuristically, where the parameter $a$ is set such that $\mathcal{R}(r) \in (-1, 1)$.
The preprocessing of the bare value $r$ is a necessary step in order to make the planar limit accessible numerically as $r \rightarrow \infty$.
The mapping via $R(r)$ is hence chosen to have finite codomain and to saturate for large absolute values of $r$, while sensitivity is retained near the origin, where curvature effects are expected to be most pronounced.

As an additional preprocessing measure, we transform the values of the discretized density profile by multiplication with the weight function $f(r, r') = r' \exp\left( - |r/\sigma - r'/\sigma| (r_f / r)^2 \right) / r$, where we set the length scale $r_f = 0.01 \sigma$ heuristically.
While such a preprocessing step is not strictly necessary (the density values are finite everywhere and hence constitute suitable input quantities in principle), this transformation of the density window helps to account for the considered geometry:
In the spherical setup, density shells near the origin correspond to a smaller volume than shells at larger distance.
To compensate for this effect, it proves to be beneficial to apply a geometric weight factor to the radially resolved density window.
Multiplying a density window with $f(r, r')$ prior to input into the neural network is a practical means to carry out such a geometric correction.
We have checked that leaving aside this preprocessing transformation, i.e.\ setting $f(r, r') = 1$, still allows for training of the neural metadensity functional, but leads to noticeably decreased quality.
The chosen form of $f(r, r')$ is currently motivated by geometrical considerations and has been determined concretely via numerical experiments.
Applications to different fluid types may necessitate further systematic fine-tuning in order to optimize the resulting quality of the neural functional.

\subsection{Pair correlation functions}
\label{SECPairCorrelationFunctions}

\subsubsection{Two-body direct correlation functional in spherical geometry}

We next lay out consequences of the geometrical setup that arise on the pair correlation level.
While we focus in this Section on general identities in the inhomogeneous fluid, we will describe specializations for the important case of bulk fluids in Sec.~\ref{SECPairCorrelationsBulk} as well as a reduction of the inhomogeneous Ornstein-Zernike equation to spherical geometry in Sec.~\ref{SECPairCorrelationsOZ}.
The relations we derive below become pertinent in particular when investigating the pair correlation structure via automatic differentiation \cite{Sammuller2023NeuralFunctionalTheory,Sammuller2024WhyNeuralFunctionals,Kampa2025MetadensityFunctionalTheory,Kampa2026MetadensityFunctionalLearning}.
We emphasize that all subsequent identities apply generically when basing the analysis on a one-body direct correlation functional $c_1(r; [\rho])$ that is resolved only along the radial coordinate $r$, which hence requires careful distinction of (implied) spherical averaging.
In particular, one needs to distinguish the response upon variations at positions $\vec{r}'$ against variations across entire shells $r'$, which we indicate in our notation by using vectorial and scalar arguments, respectively.

Results for the pair correlation structure will be scrutinized in Sec.~\ref{SECPairCorrelationStructure} for the hard sphere fluid based on $c_1(r; [\rho])$ expressed via fundamental measure theory, or alternatively represented as a neural hard sphere functional, as well as using the spherical neural metadensity functional $c_1(r; [\rho, \beta\phi])$.
To express this generality, we drop the metadensity dependence on the scaled pair potential $\beta\phi(r)$ in the notation.

We start with the standard functional derivative definition \eqref{EQc2Fromc1} of the fully resolved two-body direct correlation functional $c_2(\rv, \rv'; [\rho])$ and consider this quantity in spherically symmetric systems.
It is thereby natural to define the radially reduced two-body direct correlation functional,
\begin{equation}
  \label{EQsphericalc2FunctionalRadiallyReduced}
    \sph{c}_2(r,r';[\rho]) = \sph{c}_2(|\rv|,r';[\rho]) = \int d\rv' \delta(r'-|\rv'|) c_2(\rv,\rv';[\rho]),
\end{equation}
where the Dirac distribution $\delta(\cdot)$ constrains the position integral to a shell with constant radius $r'$.
The spherically reduced functional $\sph{c}_2(r,r';[\rho])$ depends solely on radial distances $r=|\rv|$ and $r'=|\rv'|$ as any dependence on angular degrees of freedom (between the vectors $\rv$ and $\rv'$) is integrated over by construction.

It is useful to also normalize Eq.~\eqref{EQsphericalc2FunctionalRadiallyReduced} with respect to the area of the primed spherical shell and thus define the spherically averaged two-body direct correlation functional as follows:
\begin{equation}
  \label{EQsphericalc2bar}
  \sm{c}_2(r,r';[\rho]) = \frac{\sph{c}_2(r,r';[\rho])}{4\pi r'^2}.
\end{equation}
For disambiguation to $c_2(\rv, \rv'; [\rho])$, we denote the implied geometric averaging over the sphere of radius $r'$ by the overbar.
The relationship with the bulk two-body direct correlation function $c_2^b(r)$ is described below.

The general two-body exchange symmetry of the position arguments, $c_2(\rv,\rv';[\rho])= c_2(\rv',\rv;[\rho])$ is passed down to
\begin{equation}
  \label{EQsphericalc2barExchangeSymmetry}
  \sm{c}_2(r,r';[\rho]) = \sm{c}_2(r',r;[\rho]).
\end{equation}
In contrast, the exchange symmetry is violated trivially by $\sph{c}_2(r, r'; [\rho])$.
When re-writing both sides of the above identity via Eq.~\eqref{EQsphericalc2bar} then a conversion factor occurs,
\begin{equation}
  \label{EQsphericalc2NotSymmetric}
  \sph{c}_2(r,r';[\rho]) = (r/r')^2 \sph{c}_2(r',r;[\rho]),
\end{equation}
which can alternatively be expressed as $(r'/r)\sph{c}_2(r,r';[\rho]) = (r/r') \sph{c}_2(r',r;[\rho])$.
Despite the benefit of the exchange symmetry \eqref{EQsphericalc2barExchangeSymmetry}, it is rather the form \eqref{EQsphericalc2FunctionalRadiallyReduced} that follows naturally from radially reduced functional differentiation,
\begin{equation}
  \label{EQsphericalc2FunctionalDerivative}
  \sph{c}_2(r,r';[\rho]) = \frac{\delta c_1(r;[\rho])}{\delta\rho(r')},
\end{equation}
which is straightforward to implement numerically via automatic differentiation.
Note that the numerical continuation of $c_1(r;[\rho])$ to negative values of $r$ implies to evaluate
\begin{equation}
  \label{EQsphericalc2Symmetrized}
  \sph{c}_2(r,r';[\rho]) \leftarrow \sph{c}_2(r,r';[\rho]) + \sph{c}_2(r,-r';[\rho])],
\end{equation}
when the density window $[r-r_w,r+r_w]$ overlaps with the origin, i.e.\ $r-r_w < -r'$.
This is accounted for when differentiating the symmetrized version \eqref{EQsphericalc1Symmetrized}.

\subsubsection{Specialization to bulk fluid}
\label{SECPairCorrelationsBulk}

We consider the bulk fluid with spatially constant density $\rho(r) = \rho_b$ in the following and lay out relations to the standard bulk pair direct correlation function $c_2^b(r)$.
Its spatial Fourier transform is denoted by $\tilde c_2^b(q) = 4\pi\int_0^\infty dr'' r'' c_2^b(r'') \sin(qr'')/q$.
The spherically averaged two-body direct correlation functional \eqref{EQsphericalc2bar} then satisfies
\begin{align}
  \label{EQsphericalc2FourierIntegral}
  \sm{c}_2(r,r';\rho_b) &= \frac{1}{2\pi^2}\int_0^\infty dq q^2 \tilde c_2^b(q) \frac{\sin(qr)}{qr}\frac{\sin(qr')}{qr'}\\
  \label{EQsphericalc2RealSpaceIntegral}
                                   &= \frac{1}{2} \int_{-1}^1 du c_2^b\big(\sqrt{r^2+r'^2-2 u r r'}\big)\\
  \label{EQsphericalc2AsPositionIntegral}
                                   &= \frac{1}{2rr'} \int_{|r-r'|}^{r+r'} dr'' r'' c_2^b(r''),
\end{align}
where Eq.~\eqref{EQsphericalc2FourierIntegral} follows from Fourier space considerations and the forms \eqref{EQsphericalc2RealSpaceIntegral} and \eqref{EQsphericalc2AsPositionIntegral} can be justified by (real space) geometry.
The respective bulk limit of Eqs.~\eqref{EQsphericalc2FunctionalRadiallyReduced} and \eqref{EQsphericalc2bar} is denoted by $\sph{c}_2(r, r'; \rho_b)$ and $\sm{c}_2(r, r'; \rho_b)$.
We recall that $\sph{c}_2(r, r';\rho_b)$ can be obtained in practice by automatic differentiation of $c_1(r; [\rho])$ at spatially constant bulk density $\rho(r) = \rho_b$, and that $\sm{c}_2(r, r';\rho_b)$ follows by normalization according to Eq.~\eqref{EQsphericalc2bar}.
We give a proof of the equivalence of expressions \eqref{EQsphericalc2FourierIntegral}--\eqref{EQsphericalc2AsPositionIntegral} in Appendix~\ref{appendix:c2spherical} along with further identities that relate $\sm{c}_2(r, r';\rho_b)$ to the standard pair direct correlation function $c_2^b(r)$.

The planar limit is recovered for $r, r' \rightarrow \infty$ whereby the difference of both position arguments becomes the planar coordinate $x = |r - r'|$.
Carrying out this limit leads to the well-known \cite{Sammuller2023NeuralFunctionalTheory,Sammuller2025NeuralDensityFunctional} expression
\begin{equation}
  \label{EQc2planar}
  \bar{c}_2(x; \rho_b) = 2 \pi \int_x^\infty d r'' r'' c_2^b(r)
\end{equation}
for the planar two-body direct correlation function, which is directly accessible with neural functional methods \cite{Sammuller2023NeuralFunctionalTheory,Sammuller2025NeuralDensityFunctional}.
We have thereby re-applied the normalization factor $4\pi r^2$ in Eq.~\eqref{EQsphericalc2AsPositionIntegral} to arrive at Eq.~\eqref{EQc2planar}.
The inverse transform from planar to spherical geometry is then given by
\begin{equation}
  \label{EQc2planarInverse}
  c_2^b(r) = - \frac{1}{2\pi r} \left. \frac{\partial \bar{c}_2(x;\rho_b)}{\partial x}\right|_{x=r},
\end{equation}
where one needs to bear in mind caveats regarding numerical differentiation and division by $r$.

One can recover $\sph{c}_2(r, r';\rho_b)$ in the bulk fluid from the planar two-body direct correlation functional $\bar{c}_2(x;\rho_b)$.
For this, we insert Eq.~\eqref{EQc2planarInverse} into Eq.~\eqref{EQsphericalc2AsPositionIntegral} and use Eq.~\eqref{EQsphericalc2bar} to obtain
\begin{equation}
  \label{EQc2sphericalBulkFromPlanar}
  \sph{c}_2(r, r'; \rho_b) = \frac{r'}{r} \left( \bar{c}_2(x=|r - r'|;\rho_b) - \bar{c}_2(x=r + r';\rho_b) \right).
\end{equation}
Equation~\eqref{EQc2sphericalBulkFromPlanar} provides a useful consistency relation for a spherical (neural) functional, as the right-hand side features explicitly the two-body direct correlation function $\bar{c}_2(x)$ in the planar limit.
We have checked that Eq.~\eqref{EQc2sphericalBulkFromPlanar} is satisfied for the case of the neural metadensity functional investigated in Sec.~\ref{SECResults}.

The above relations demonstrate that the two-body direct correlation function reduced to spherical geometry, as given either by $\sph{c}_2(r,r')$ or $\sm{c}_2(r, r')$, allows one to recover consistently the bulk pair correlation structure, thereby encompassing also the planar limit that arises for large radial distance to the origin.
As $\sph{c}_2(r,r';[\rho])$ results immediately from functional differentiation of $c_1(r; [\rho])$, cf.\ Eq.~\eqref{EQsphericalc2FunctionalDerivative}, these relations prove to be pertinent in subsequent applications of the neural metadensity functional, see Sec.~\ref{SECPairCorrelationStructure}.

\subsubsection{Spherically averaged Ornstein-Zernike equation}
\label{SECPairCorrelationsOZ}

We now return to the inhomogeneous fluid and lay out the relation of spherically averaged direct and total correlation functions via reduction of the inhomogeneous Ornstein-Zernike equation~\eqref{EQOZinhom} to spherical symmetry.
Doing so is feasible by integration of Eq.~\eqref{EQOZinhom} over spherical shells $r = |\rv|$ and $r' = |\rv|$.
Recalling the definition \eqref{EQsphericalc2bar} of the spherically averaged direct correlation function, we define $h_2(r, r')$ and $\sm{h}_2(r, r')$ in the same fashion starting from the fully resolved quantity $h_2(\rv, \rv')$.
Then, the spherically averaged Ornstein-Zernike equation attains the simple form
\begin{equation}
  \label{EQOZspherical}
  \begin{split}
    \sm{h}_2(r, r') &= \sm{c}_2(r, r')\\
                   &\quad+ 4\pi \int dr'' r''^2 \sm{h}_2(r, r'') \rho(r'') \sm{c}_2(r'', r'),
  \end{split}
\end{equation}
hence keeping its original integral structure \eqref{EQOZinhom} intact (note the additional normalization factor $4\pi r''^2$ in the second term on the right hand side).
An alternative derivation of Eq.~\eqref{EQOZspherical} proceeds via Legendre (polynomial) transformation of Eq.~\eqref{EQOZinhom}, which yields a hierarchy of integral equations for the associated Legendre transforms of $c_2(\rv, \rv')$ and $h_2(\rv, \rv')$ \cite{Attard1989SphericallyInhomogeneousFluids,Tschopp2021FundamentalMeasureTheory}.
Equation~\eqref{EQOZspherical} then arises as the lowest-order member of the hierarchy, which crucially decouples from all higher angular components.
Due to the numerical accessibility of $\sph{c}_2(r, r'; [\rho])$ via automatic differentiation, cf.\ Eq.~\eqref{EQsphericalc2FunctionalDerivative}, the solution of Eq.~\eqref{EQOZspherical} constitutes a practically feasible route to the spherically averaged total correlation function.
We have all information to define the spherically averaged two-body density
\begin{equation}
  \label{EQbarrho2}
  \sm{\rho}_2(r, r') = \rho(r) \rho(r') (\sm{h}_2(r, r') + 1)
\end{equation}
to quantify spherically averaged density-density correlations, as will be scrutinized for concrete applications below.

\section{Results}
\label{SECResults}

\subsection{Training of spherical neural functionals}
\label{SECTraining}

\begin{figure}[tb]
  \includegraphics[width=\columnwidth]{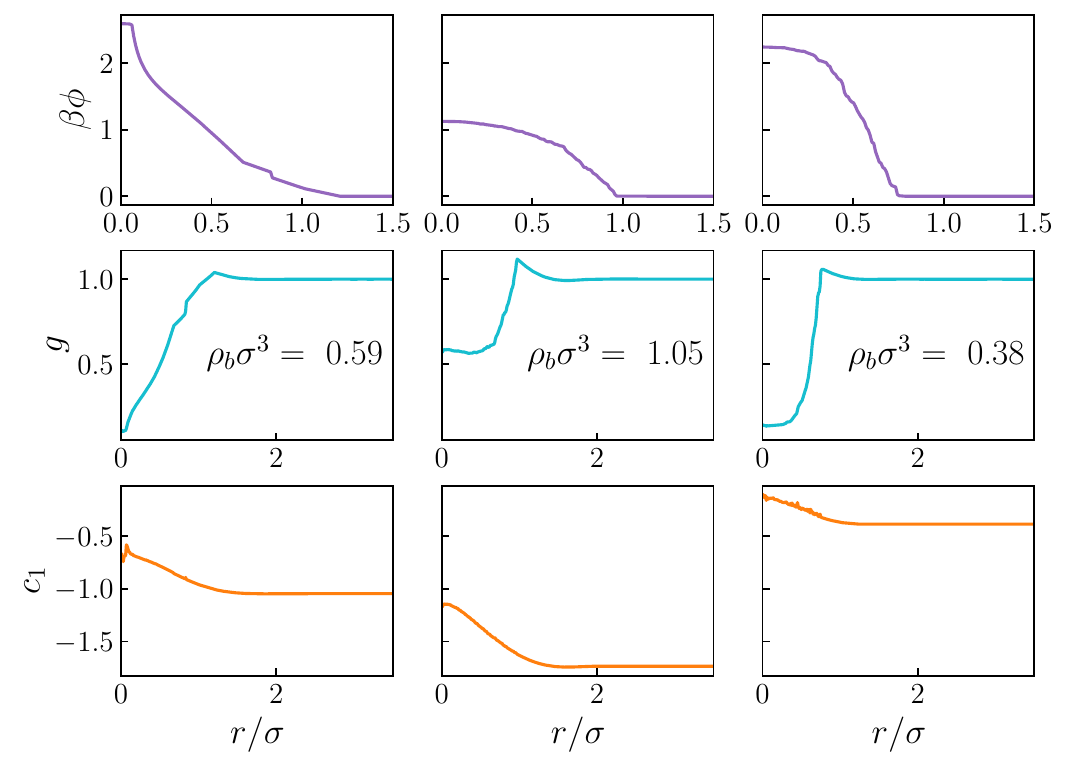}
  \caption{
    Illustration of three exemplary test particle results (columns) used for training of the neural metadensity functional in spherical geometry.
    Randomized forms of the scaled pair potential $\beta \phi(r)$ (first row) yield corresponding radial distribution functions $g(r)$ (middle row) and bulk densities (not shown), which correspond to one-body correlation functions $c_1(r)$ (bottom row) according to Eq.~\eqref{EQsphericalTestParticleTrainingData}, whereby $r / \sigma$ denotes the scaled distance from the origin, along which all profiles are resolved.
  }
\label{FIGtraindata}
\end{figure}

We use $\sim 3000$ bulk grand canonical Monte Carlo simulations in cubic boxes of lateral length $L = 15 \sigma$, where the chemical potential is chosen uniformly in $\mu \in [-3, 2]$.
The generation of repulsive pair potentials follows the randomization procedure described in Ref.~\cite{Kampa2025MetadensityFunctionalTheory} and we choose $r_c = 1.5 \sigma$ as truncation distance such that $\phi(r) = 0$ for $r > r_c$.
We measure radial distribution functions $g_k(r)$ and bulk densities $\rho_{b,k}$ in each simulation $k$ and take these effective test particle data as input for the calculation of $c_{1,k}(r)$ via Eq.~\eqref{EQsphericalTestParticleTrainingData}.
Exemplary test particle results are shown in Fig.~\ref{FIGtraindata}.
In addition, we supplement the training data set by $\sim 1500$ simulations in planar geometry, where a randomized external potential $V_{\rmext,k}(x)$ is imposed, following Refs.~\cite{Sammuller2023NeuralFunctionalTheory,Kampa2025MetadensityFunctionalTheory}.
The resulting inhomogeneous density profiles yield $c_{1,k}(x)$ via Eq.~\eqref{EQplanarTrainingData}.

Our neural network consists of five fully connected hidden layers with 512, 256, 128, 64, and 32 nodes, respectively.
The input layer is divided into three channels: 401 nodes accept the local density profile $\rho(r')|_{|r - r'| < r_w}$ within a cutoff distance $r_w = 2.0 \sigma$.
150 nodes receive the scaled pair potential $\beta \phi(r)$, which we recall vanishes beyond a truncation distance $r_c = 1.5 \sigma$.
Both functions are represented by discretization on an equidistant grid with discretization interval $\Delta r = 0.01 \sigma$.
An additional input node encodes curvature information by supplying it with the value of $\mathcal{R}(r)$ for the considered radial distance $r$.
Training proceeds via optimization based on the matching conditions~\eqref{EQsphericalTestParticleMatching}, \eqref{EQsphericalBulkMatching}, and \eqref{EQplanarMatching} using standard batched training in 250 epochs with the Adam optimization routine for backpropagation of errors; we recall the overview given in Sec.~\ref{SECSphericalMetadensityFunctionalLearning}.
Note that the offset $r_\infty$ in Eq.~\eqref{EQplanarMatching} for recovering the planar limit is implemented by setting the curvature input $\mathcal{R}(r = \pm \infty) = \pm 1$, corresponding to infinitely large distance from the origin, in our local learning scheme.

As an additional means of comparison, we have considered the hard sphere fluid and trained a second neural functional, which does \emph{not} include any metadensity functional dependence.
In contrast to the test particle data used for the previous metadensity functional training, the training data for the hard sphere functional consist of simulations in spherically inhomogeneous environments, which are processed on the basis of Eqs.~\eqref{EQinhomTrainingData} and \eqref{EQinhomMatching}.
In total, we use $\sim 2500$ individual simulation results corresponding to confinement in hard spherical shells and cavities with randomized (inner and outer) radius and with randomized chemical potential $\mu \in [-3, 2]$.
These data crucially include confinement in very small cavities and narrow shells.
Additionally, we provide reference data from $\sim 900$ hard sphere simulations in planar geometry.
The normalizing function for the curvature input has been simplified to $\mathcal{R}(r) = \tanh(a r)$ with heuristically chosen inverse length scale $a = 0.3 / \sigma$.

\subsection{Prediction of inhomogeneous density profiles}
\label{SECPrediction}

\begin{figure*}[tb]
  \includegraphics[width=\textwidth]{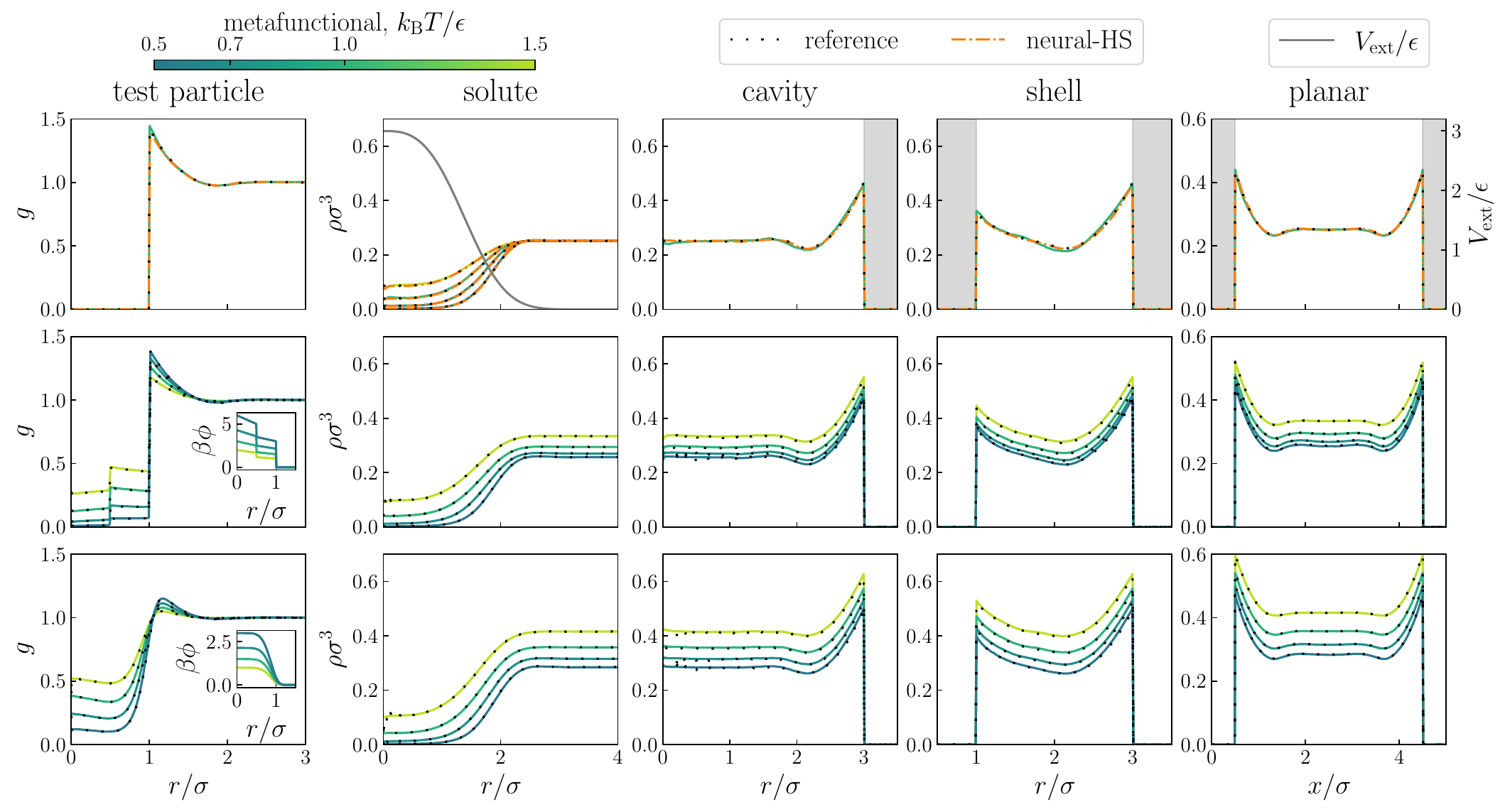}
  \caption{
    Neural metadensity functional predictions for density profiles (solid lines) in selected inhomogeneous external potentials $V_\rmext(r)$ (grey lines).
    We consider a test particle fixed at the origin (first column), a penetrable solute at the origin (second column), a spherical cavity of radius $3\sigma$ (third column), confinement in a hard spherical shell (fourth column), and confinement between planar hard walls (fifth column).
    The scaled chemical potential of all systems is $\beta \mu = 0$.
    The interaction potentials encompass hard spheres at fixed temperature (first row) as well as a two-step interaction potential (second row) and penetrable spheres (third row) at multiple scaled temperature values $\kB T / \epsilon$ with energy unit $\epsilon$, as indicated by the colorbar; the insets in the first column depict the respective form of the scaled pair potential $\beta \phi(r)$.
    We compare the hard sphere predictions of the metadensity functional to results from a separately trained neural hard sphere functional (``neural HS'', orange lines), see text, as well as to Rosenfeld's fundamental measure density functional theory \cite{Rosenfeld1989FreeEnergyModel,Roth2010FundamentalMeasureTheory} (black dotted lines).
    In the second and third row, the metadensity functional predictions are confirmed by comparison to simulation data (black dotted lines).
    Note that all metadensity predictions stem from the same neural functional.
  }
\label{FIGpredictions}
\end{figure*}

In Fig.~\ref{FIGpredictions}, we depict results for equilibrium density profiles obtained with the trained neural metadensity functional in various inhomogeneous settings and for different choices of the scaled pair potential $\beta \phi(r)$.
In practice, we rearrange the radially resolved Euler-Lagrage equation~\eqref{EQinhomTrainingData} to
\begin{equation}
  \label{EQPicard}
  \rho(r) = \exp\left( -\beta \left(V_\rmext(r) - \mu\right) + c_1(r;[\rho,\beta\phi]) \right).
\end{equation}
We use mixed Picard iteration of Eq.~\eqref{EQPicard} to determine $\rho(r)$ self-consistently for the desired choices of thermodynamic parameters $\beta$, $\mu$ and external potential $V_\rmext(r)$, thereby evaluating $c_1(r; [\rho, \beta\phi])$ with the neural metadensity functional for the specific scaled pair potential $\beta \phi(r)$ under consideration.

We consider hard spheres, a repulsive pair potential with two plateaus, and a soft penetrable potential, see the insets in Fig.~\ref{FIGpredictions} for illustrations.
The inhomogeneous situations that we investigate encompass a test particle setup where $V_{\rmext}(r) = \phi(r)$, a solute at the origin with Gaussian (soft) repulsion, confinement both in a hard spherical cavity and shell, as well as confinement between two parallel hard walls.

We recall that planar geometry is recovered by setting the neural network input $\mathcal{R}(r = \infty) = 1$.
Due to the symmetric continuation of the density profile to negative values of $r$, setting $\mathcal{R}(r = -\infty) = -1$ also yields the planar limit; in applications both values can hence be utilized and averaged.
Similar symmetrization is possible by utilizing mirror symmetry in planar geometry and evaluating the neural functional for flipped density input.

For the penetrable pair potentials, we compare the metadensity functional predictions to simulation data from grand canonical Monte Carlo simulations and find excellent agreement.
Predictions of the metadensity functional for the case of hard spheres are compared to fundamental measure theory in Rosenfeld's formulation \cite{Rosenfeld1989FreeEnergyModel,Roth2010FundamentalMeasureTheory}.
All investigated metadensity functional results agree closely with the respective reference.

\subsection{Henderson inversion}
\label{SECHenderson}

\begin{figure*}[tb]
  \includegraphics[width=\textwidth]{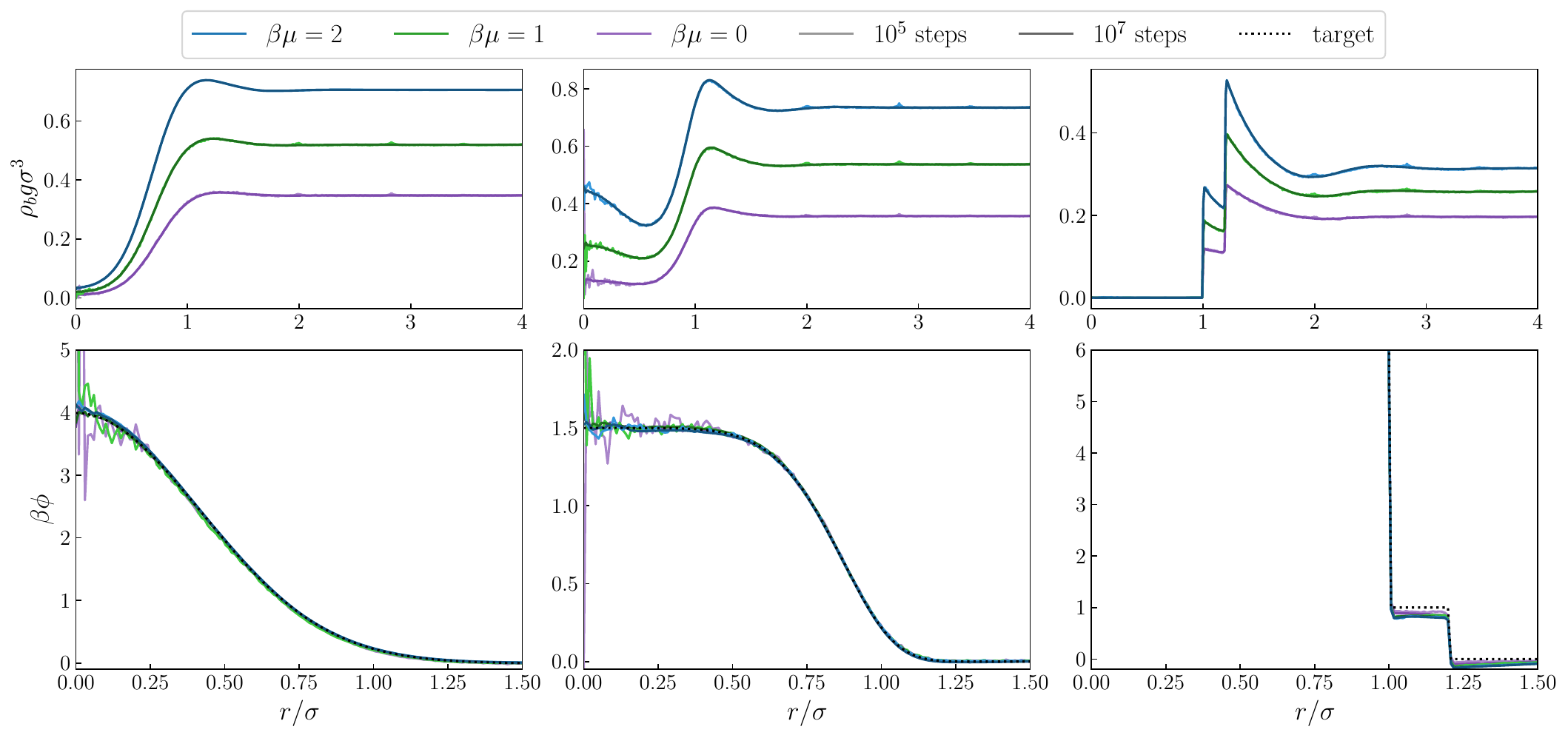}
  \caption{
    Henderson inversion from given radial distribution functions $g(r)$ at bulk densities $\rho_b$ to the corresponding scaled pair potentials $\beta \phi(r)$.
    As input, we take $\rho_b g(r)$ (first row) obtained from bulk fluid simulations at $\kB T / \epsilon = 1.0$ and at different values of the scaled chemical potential $\beta \mu = 0, 1, 2$ (as indicated).
    The neural metadensity functional then allows one to recover the corresponding scaled pair potentials (lower rows), as shown for three different representative choices (each column, respectively).
    The inversion is successful in all considered cases, as can be verified against the known forms of $\beta \phi(r)$ that were employed in the simulations (black dotted lines).
    This continues to hold when $g(r)$ shows increased noise artifacts resulting from simulations with fewer samples (desaturated lines).
  }
\label{FIGHenderson}
\end{figure*}

The explicit access to the \emph{functional} dependence on the pair potential allows flexible use in inverse design tasks.
We here consider Henderson inversion in spherical three-dimensional geometry, where the aim is to find the unique \cite{Henderson1974UniquenessTheoremFluid} pair potential that gives rise to a prescribed form of the radial distribution function $g(r)$ at given bulk density $\rho_b$ and for prescribed temperature $T$.
While this task has been demonstrated on the basis of neural metadensity functional theory in one-dimensional geometry \cite{Kampa2025MetadensityFunctionalTheory}, the application to three-dimensional fluids requires the ability to resolve spherical inhomogeneities as occur in the employed test particle setup (see below).
We emphasize that planar symmetry is hence insufficient for immediate application of the inversion scheme laid out below.
To investigate the feasibility of Henderson inversion in three dimensions, we acquire data for $g(r)$ and $\rho_b$ in bulk fluid simulations for three exemplary choices of the scaled pair potential $\beta \phi(r)$.
Despite the pair potential being hence known explicitly, we deliberately seek to \emph{recover} $\beta \phi(r)$ by solely making use of the neural metadensity functional, as applied to the simulation results for $g(r)$ and $\rho_b$.

We follow Ref.~\cite{Kampa2025MetadensityFunctionalTheory} to conceive a practical scheme and first reexpress the scaled chemical potential as
\begin{equation}
  \label{EQmuBulk}
  \beta \mu = \ln\rho_b - c_1(r; \rho_b, [\beta\phi])
\end{equation}
by considering the Euler-Lagrange equation in the bulk limit for vanishing $V_\rmext(r)$; note that the functional is evaluated at constant bulk density $\rho_b$ and that the chemical potential must be spatially constant in equilibrium.
Rearrangement of the test particle Eq.~\eqref{EQsphericalTestParticleTrainingData} using Eq.~\eqref{EQmuBulk} then yields
\begin{equation}
  \label{EQHenderson}
  \begin{split}
    \beta \phi(r) &= c_1(r; [\rho_b g(r), \beta\phi]) - \ln(\rho_b g(r))\\
                  &\quad- c_1(r; \rho_b, [\beta\phi]) + \ln\rho_b,
  \end{split}
\end{equation}
which now constitutes an implicit equation for $\beta\phi(r)$, in contrast to Eq.~\eqref{EQPicard}, where the functional argument $\beta \phi(r)$ remained fixed.
In particular, all quantities on the right hand side of Eq.~\eqref{EQHenderson} are accessible in practice, without the need to carry out simulations, as they are either prescribed or readily evaluated using the neural metadensity functional.
A solution for $\beta \phi(r)$ is obtained by self-consistent mixed Picard iteration, which we find to converge rapidly in all investigated scenarios.

In Fig.~\ref{FIGHenderson}, we apply this scheme to recover the three considered pair potentials solely from the given forms of $g(r)$ and values of $\rho_b$ and $T$ that originate from sampling in grand canonical Monte Carlo simulations.
We find reliable results for all cases and for all thermodynamic state points considered.
In particular, the inversion remains successful for (slightly) noisy results of $g(r)$, as follows from using simulation data with fewer samples.
Small discrepancies of the recovered pair potentials $\phi(r)$ to the (known) ground truth are visible particularly near the origin $r = 0$ and for the case of the (discontinuous) square-shoulder potential, where the energy plateau at $1 < r/\sigma < 1.2$ is slightly underestimated, cf.\ the third column in Fig.~\ref{FIGHenderson}.
Leaving these small numerical deviations aside, we find Henderson inversion via fixed-point iteration \eqref{EQHenderson} with the neural metadensity functional to be a robust tool for the recovery of the present short-ranged pair potentials.
These observations prompt further conceptual work to scrutinize the underlying physical mechanism, given the difficulty of addressing this task with conventional computational methods.

\subsection{Pair correlation structure}
\label{SECPairCorrelationStructure}

\begin{figure}[tb]
  \includegraphics[width=\columnwidth]{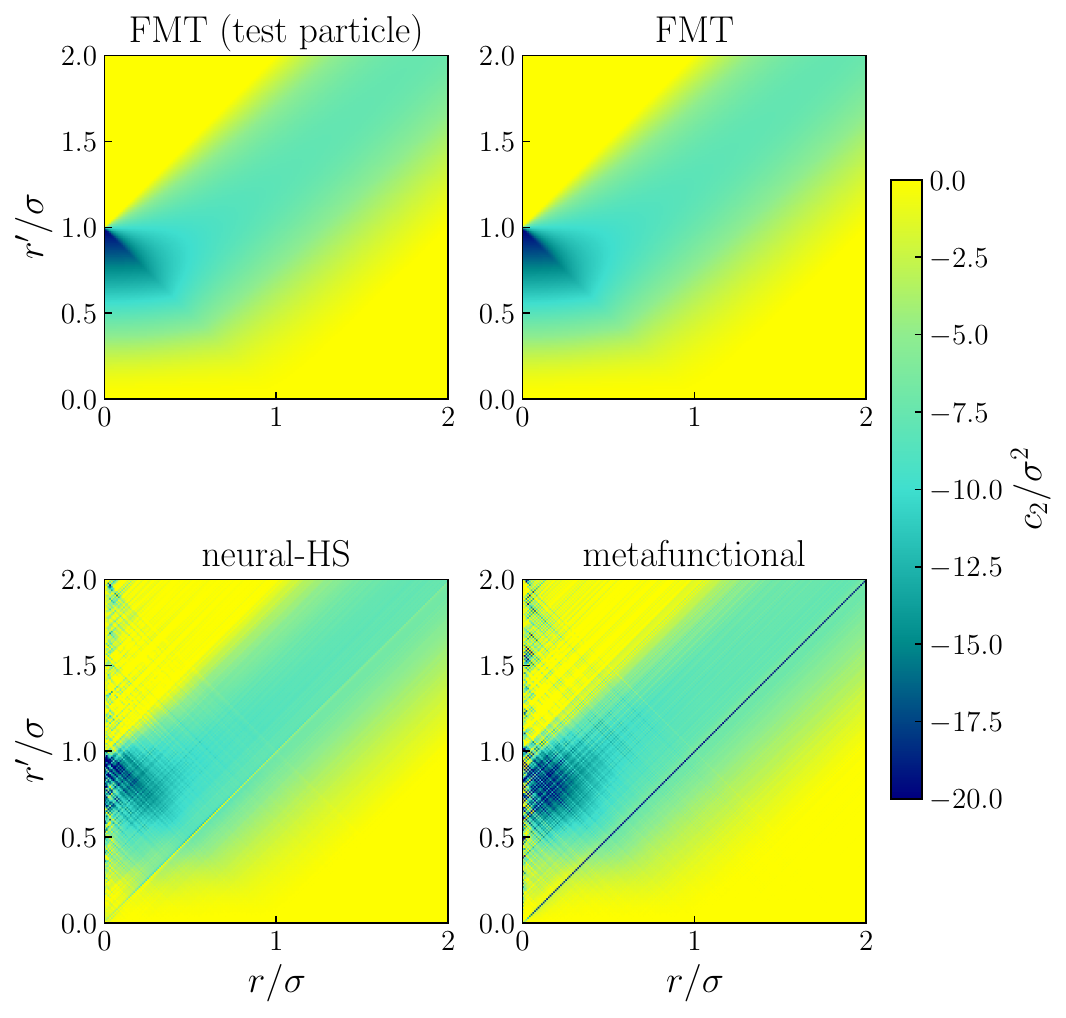}
  \caption{
    Radially resolved two-body direct correlation function $\sph{c}_2(r, r')$ for the bulk hard sphere fluid at scaled chemical potential $\beta \mu = 1$, corresponding to a bulk density $\rho_b \sigma^3 \approx 0.34$.
    Depicted are results from automatic differentiation of the neural metadensity functional (``metafunctional'') and from automatic differentiation of a separately trained functional that is purely applicable to the hard sphere fluid (``neural-HS'').
    For comparison, we show results from fundamental measure theory, whereby we consider again simple automatic differentiation of the analytic fundamental measure functional (``FMT'').
    As an additional consistency check (``FMT (test particle)''), we use fundamental measure theory in a spherical test particle setup to compute $g(r)$, which is converted via the Ornstein-Zernike equation to $c_2^b(r)$, projected to the planar representation $\bar{c}_2(x)$ via Eq.~\eqref{EQc2planar}, and ultimately transformed to $\sph{c}_2(r, r')$ via Eq.~\eqref{EQc2sphericalBulkFromPlanar}.
  }
\label{FIGc2homogeneous}
\end{figure}

\begin{figure}[tb]
  \includegraphics[width=\columnwidth]{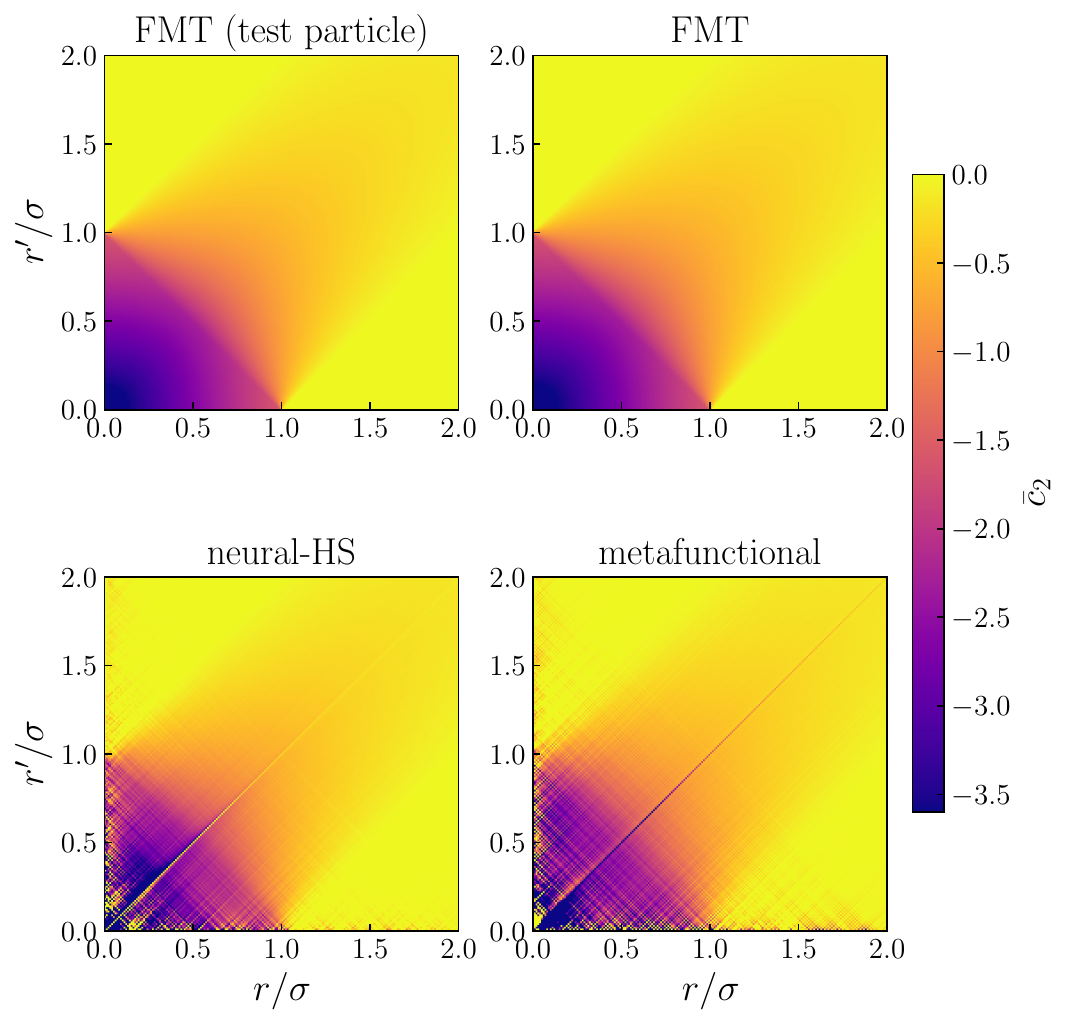}
  \caption{
    Spherically averaged two-body direct correlation function $\sm{c}_2(r, r')$ for the bulk hard sphere fluid.
    The results for $\sph{c}_2(r, r')$ as shown in Fig.~\ref{FIGc2homogeneous} are thereby symmetrized according to Eq.~\eqref{EQsphericalc2Symmetrized}.
  }
\label{FIGringc2homogeneous}
\end{figure}

\begin{figure}[tb]
  \includegraphics[width=\columnwidth]{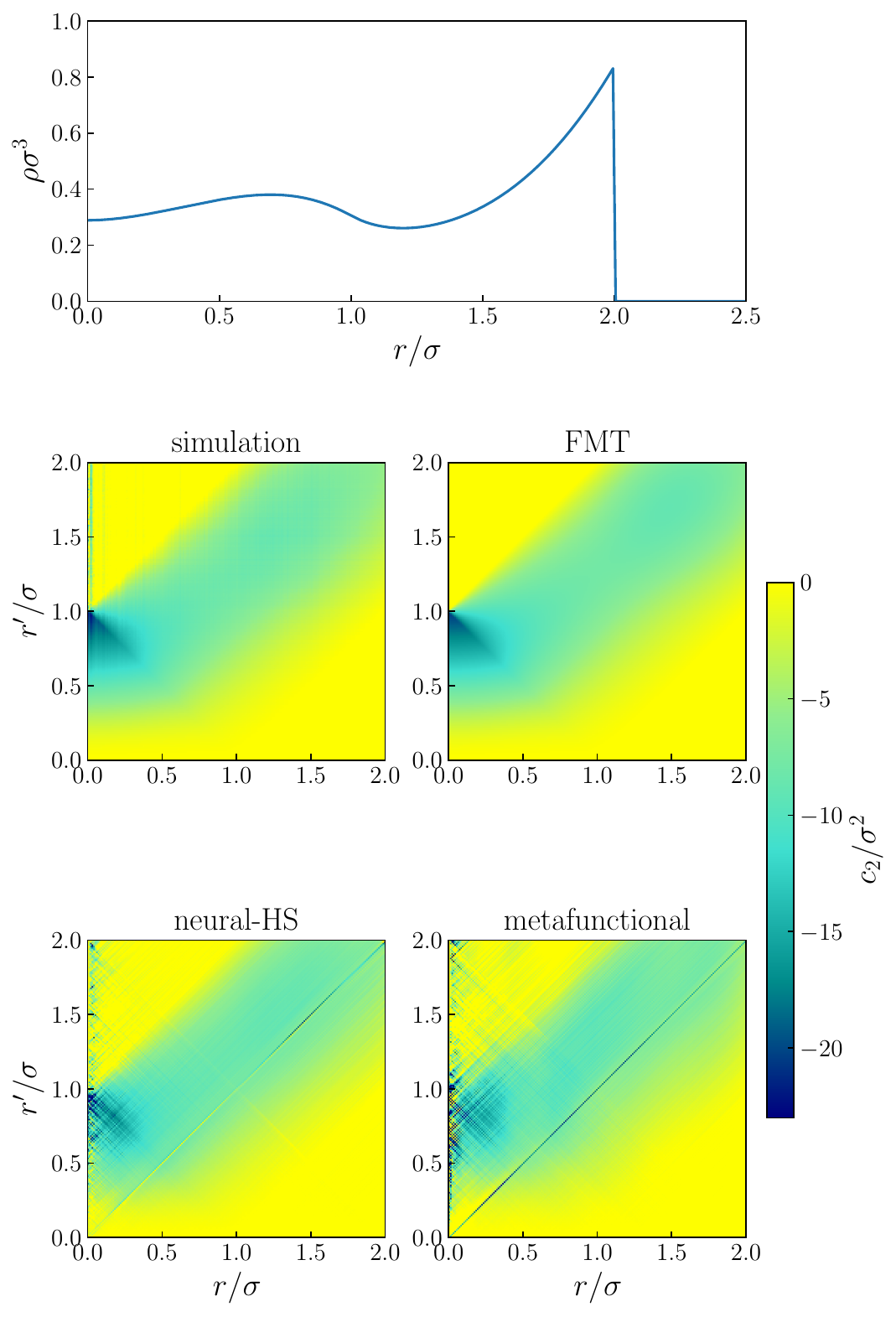}
  \caption{
    Radially resolved two-body direct correlation function $\sph{c}_2(r, r')$ for the hard sphere fluid confined in a spherical cavity of radius $2\sigma$.
    The corresponding density profile is depicted in the upper panel.
    We investigate again different routes, whereby ``FMT'', ``neural-HS'', and ``metafunctional'' correspond to those shown in Fig.~\ref{FIGc2homogeneous}; see the caption and the text for further details.
    As reference, we also show $\sph{c}_2(r, r')$ obtained from simulations; see text.
  }
\label{FIGc2cavity}
\end{figure}

\begin{figure}[tb]
  \includegraphics[width=\columnwidth]{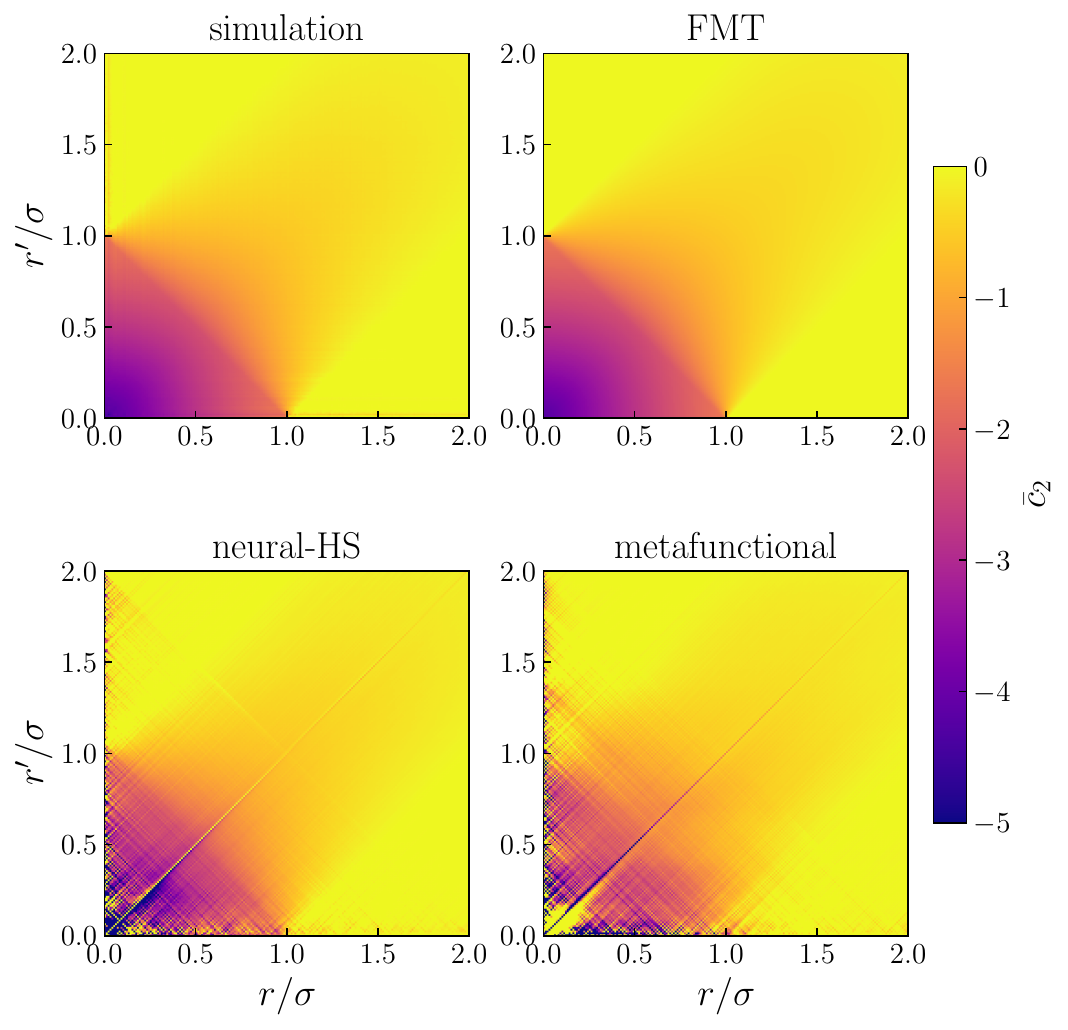}
  \caption{
    Spherically averaged two-body direct correlation function $\sm{c}_2(r, r')$ for the hard sphere fluid confined in a spherical cavity from the same routes as investigated in Fig.~\ref{FIGc2cavity}.
  }
\label{FIGringc2cavity}
\end{figure}

In the following, we consider the application of automatic differentation \cite{Baydin2018AutomaticDifferentiationMachine,Stierle2024ClassicalDensityFunctional} in order to investigate the pair correlation structure in both homogeneous and spherically heterogeneous conditions.
Recall that simple application of a density functional derivative immediately yields the radially resolved quantity $\sph{c}_2(r, r'; [\rho, \beta\phi])$, cf.\ Eq.~\eqref{EQsphericalc2FunctionalDerivative}, where (single) spherical averaging is implied according to Eq.~\eqref{EQsphericalc2FunctionalRadiallyReduced}.
For the bulk hard sphere fluid, we depict results for $\sph{c}_2(r, r'; [\rho, \beta\phi])$ obtained by straightforward automatic differentiation of the neural metadensity functional in Fig.~\ref{FIGc2homogeneous}.
We compare these to results from automatic differentiation of both a neural hard sphere functional (without dependence on the pair potential) and of the analytical Rosenfeld functional.
The latter fundamental measure theory functional is used in a further route, which yields $\sph{c}_2(r, r')$ without using automatic differentiation.
Instead, we calculate $g(r)$ via test particle minimization, obtain $c_2^b(r)$ via Ornstein-Zernike inversion, compute $\bar{c}_2(x; \rho_b)$ by numerical integration according to Eq.~\eqref{EQc2planar}, and ultimately find $\sph{c}_2(r, r'; \rho_b)$ from Eq.~\eqref{EQc2sphericalBulkFromPlanar}.
This substantially different route serves to demonstrate the consistency of the above bulk two-body relations and of our employed numerics.

The lacking symmetry of $\sph{c}_2(r, r')$ with respect to exchanging $r$ and $r'$ is notable and expected due to Eq.~\eqref{EQsphericalc2NotSymmetric}, which highlights the importance of the careful consideration of curvature effects that arise in spherical geometry, as developed in Sec.~\ref{SECPairCorrelationFunctions}.
Although $\sph{c}_2(r, r')$ is arguably the more immediate quantity to consider, as it is obtained directly from automatic differentiation of $c_1(r; [\rho])$ via Eq.~\eqref{EQsphericalc2FunctionalDerivative}, it is useful to transform the result to the spherically averaged two-body direct correlation function $\sm{c}_2(r, r')$, which compensates for the different shell volumes at $r$ and $r'$.
In Fig.~\ref{FIGringc2homogeneous}, we depict results from the same routes as in Fig.~\ref{FIGc2homogeneous}, but utilize the identity \eqref{EQsphericalc2Symmetrized} to arrive at $\sm{c}_2(r, r')$.
This quantity satisfies exchange symmetry in $r$ and $r'$, as can be gleaned from Eq.~\eqref{EQsphericalc2barExchangeSymmetry}.

For the case of inhomogeneous input, $\sph{c}_2(r, r')$ continues to be easily accessible via automatic differentiation.
In Fig.~\ref{FIGc2cavity}, we depict results corresponding to confinement of the hard sphere fluid in a hard spherial cavity of radius $2\sigma$.
For comparison, we consider results from a neural hard sphere functional without metadensity dependence as well as from fundamental measure theory.
Additionally, simulation results for $\sm{\rho}_2(r, r')$ are acquired in grand canonical Monte Carlo simulation and transformed via the spherically averaged Ornstein-Zernike equation \eqref{EQOZspherical} and Eq.~\eqref{EQsphericalc2bar} to $\sph{c}_2(r, r')$.
These serve as further reference.
The quantity $\sm{c}_2(r, r')$ is shown in Fig.~\ref{FIGringc2cavity} for the identical setup and routes, confirming the exchange symmetry in $r$ and $r'$ also for this inhomogeneous situation.
The fact that all results are accessible and agree with each other confirms the correct incorporation of the geometric setup in our methodology.
We emphasize in this regard the intricate features that are apparent near the origin, i.e.\ for small values of $r$, which originate in the spherically symmetric setup, see in particular Figs.~\ref{FIGc2homogeneous} and \ref{FIGc2cavity}.
These purely geometric effects are reproduced successfully by both of our neural functionals.
In Appendix~\ref{appendix:two-body}, we demonstrate the accessibility of inhomogeneous total pair correlation functions $\sm{h}_2(r,r')$ and two-body densities $\sm{\rho}_2(r,r')$ on the basis of the spherically averaged Ornstein-Zernike equation \eqref{EQOZspherical}.

\subsection{Test particle thermodynamics}
\label{SECTestParticleThermodynamics}

\begin{figure}[tb]
  \includegraphics[width=\columnwidth]{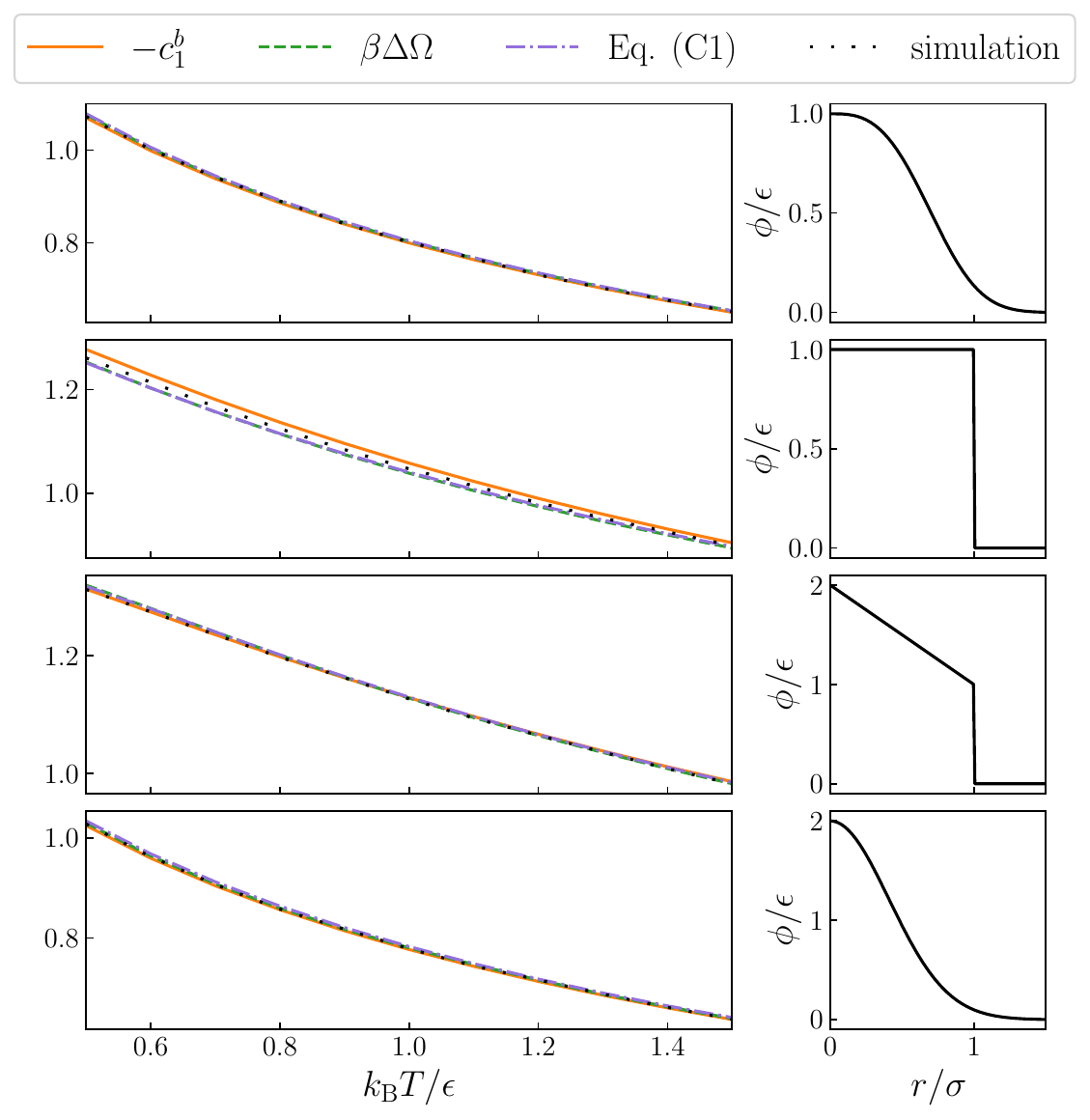}
  \caption{
    Verification of thermodynamic test particle sum rules (left hand side) on the basis of the neural metadensity functional.
    We show the scaled chemical potential $\beta \mu_\rmexc$ as obtained from three different routes: (i) mere evaluation of the neural metadensity functional according to Eq.~\eqref{EQPercusMuExc}, (ii) evaluation of the grand potential difference according to Eq.~\eqref{EQPercusMuExc} using a linear parameterization \eqref{EQfunctionalLineIntegralLinear} for both arising functional line integrals, and (iii) calculation of the alternative functional line integral according to Eq.~\eqref{EQPercusThermodynamicAsFunctionalIntegral}.
    Reference data is taken from simulations (dotted black lines).
    The test particle sum rule is investigated for four different choices (rows) of the pair potential $\phi / \epsilon$ (right hand side).
  }
\label{FIGtpsumrule}
\end{figure}

In contrast to the structural properties investigated in the previous section, we now seek to assess whether the neural metadensity functional satisfies fundamental \emph{thermodynamic} consistency relations.
Specifically, we lay out various routes toward the bulk excess chemical potential $\mu_\rmexc$, which is defined via
\begin{equation}
  \label{EQMuExc}
  \beta \mu_\rmexc = \beta\mu-\ln\rho_b = -c_1(r;\rho_b,[\beta\phi]) = -c_1^b,
\end{equation}
where $\rho_b$ indicates the bulk fluid density and where we have defined the bulk value of the one-body direct correlation functional $c_1^b$, which is trivially independent of position $r$.
Equation~\eqref{EQMuExc} serves as the first means of determining $\beta \mu_\rmexc$, as $c_1^b$ is directly accessible via evaluation of the neural metadensity functional, as shown in Fig.~\ref{FIGtpsumrule} for four different choices of the pair potential $\phi(r)$.

Following Percus, see e.g.\ Refs.~\cite{Percus1962ApproximationMethodsClassical,Gul2024UsingTestParticle,Gul2026UsingTestParticle}, the test particle excess chemical potential may equivalently be expressed as
\begin{align}
  \label{EQPercusMuExc}
  \mu_\rmexc &= \Omega[\rho_b g,\beta\phi] - \Omega_0(\rho_b,[\beta\phi]),
\end{align}
such that, for the given scaled interparticle pair potential $\beta\phi(r)$, the bulk grand potential $\Omega_0(\rho_b,[\beta\phi])$ is subtracted from the grand potential of the test particle state $\Omega[\rho_b g,\beta\phi]$.
Both values of the grand potential are determined via Eq.~\eqref{EQOmega}, whereby the excess free energy is evaluated using functional line integration \cite{Sammuller2023NeuralFunctionalTheory,Sammuller2024WhyNeuralFunctionals,Sammuller2025NeuralDensityFunctional}:
\begin{equation}
  \label{EQfunctionalLineIntegralLinear}
  \begin{split}
  F_\rmexc[\rho, \beta\phi] &= -\kB T \int d\rv \rho(\rv) \int_0^1 da c_1(\rv; [\rho_a,\beta\phi])\\
    &= -\kB T 4 \pi \int_0^\infty dr r^2 \rho(r) \int_0^1 da c_1(r; [\rho_a,\beta\phi])
  \end{split}
\end{equation}
A linear parameterization is thereby used such that $\rho_a(r) = a \rho(r)$ with $0 < a < 1$, and we perform the integration numerically, evaluating the integrand $c_1(r; [\rho_a,\beta\phi])$ with the neural metadensity functional, as is possible due to the learned radial dependence.
Importantly, the spherical geometry does not hamper the fundamental concept of functional line integration.
Results are shown in Fig.~\ref{FIGtpsumrule}, demonstrating that our neural results confirm to Eq.~\eqref{EQPercusMuExc}.
To carry out a consistency check, we have investigated an alternative functional integration method (see Appendix~\ref{appendix:testparticle}), which relies on a different parameterization of the functional line integral.
The results from this route agree numerically with the results of the former two routes described above.

\section{Conclusions}
\label{SECConclusions}

In conclusion, we have laid out a machine-learning-based scheme to obtain practical access to a neural metadensity functional that is applicable for addressing general short-ranged pairwise interparticle interactions that govern three-dimensional fluids in spherically symmetric environments.
Classical density functional theory \cite{Evans1979NatureLiquidvapourInterface,Evans1992DensityFunctionalsTheory,Hansen2013TheorySimpleLiquids} provides the mathematical footing and establishes the uniqueness of functionals that depend both on the inhomogeneous density profile $\rho(r)$ as well as on the thermally scaled pair potential $\beta \phi(r)$.
The metadensity functional concept \cite{Kampa2025MetadensityFunctionalTheory,Kampa2026MetadensityFunctionalLearning} makes the latter functional dependence operational in practice.
In this regard, we have extended previous investigations for prototypical one-dimensional fluids \cite{Kampa2025MetadensityFunctionalTheory,Kampa2026MetadensityFunctionalLearning} to the class of three-dimensional fluids in spherical geometry where particles interact with virtually arbitrary repulsive pair potentials of finite range.

Dealing with the functional dependencies in practice is feasible by employing neural metadensity functional learning \cite{Sammuller2023NeuralFunctionalTheory,Sammuller2024WhyNeuralFunctionals,Kampa2025MetadensityFunctionalTheory}, which we have specialized here to fluids in spherically symmetric external potentials.
Despite the rich inhomogeneous structure that this setup encompasses, we have shown that test particle data together with inhomogeneous data in planar symmetry is sufficient for training a neural metadensity functional.
All required quantities in the test particle setup are already available in bulk fluid simulations, such that the computational effort remains very moderate as compared to simulating explicitly general spherical inhomogeneities.
Adapting the neural network and the local learning strategy to the considered geometry allowed us to obtain a faithful representation of the underlying functional mapping of the one-body direct correlation functional $c_1(r; [\rho, \beta\phi])$, including the crucial curvature effects, with high accuracy even in regions where these are most pronounced, i.e.\ at small distances $r$ from the origin.
While we have kept the training routine as simple as possible for the present investigation, we can envisage further refinement by incorporating recently developed regularization techniques for neural density functionals \cite{Sammuller2024NeuralDensityFunctionals,Kampa2026MetadensityFunctionalLearning}.

The spherical neural metadensity functional is applicable for addressing a significant variety of physical problems.
We have demonstrated the prediction of density profiles in prescribed external environments, including test particle situations, inhomogeneous structuring in spherical cavities and shells, as well as confinement in planar (flat-wall) geometry; the latter scenario corresponds to the case of vanishing curvature in the limit of large radial separation, which is systematically incorporated in our methodology.
The challenging inverse (Henderson) problem of inferring the pair potential for given bulk density and radial distribution function has further been scrutinized.
Similar to earlier demonstrations in one-dimensional geometry \cite{Kampa2025MetadensityFunctionalTheory}, we find this inversion \cite{Henderson1974UniquenessTheoremFluid} to be feasible in the present spherical setup using the neural metadensity functional in the solution of an \emph{implicit} equation that determines the pair potential.
We attribute this theory-driven approach to be the key mechanism that facilitates successful inversion also for the three-dimensional fluids considered in this work.

The pair correlation structure in both homogeneous and spherically heterogeneous fluids is immediately and efficiently accessible via automatic functional differentiation.
We emphasize that the utilization of automatic differentiation constitutes a general mechanism, which we have applied both to neural (metadensity) functionals but also to fuctionals implemented analytically based on fundamental measure theory \cite{Rosenfeld1989FreeEnergyModel,Roth2010FundamentalMeasureTheory}.
Substantial cross-checks of radially resolved pair direct correlation functions for the case of the hard sphere fluid against results from multiple functionals and routes confirm all respective predictions that we have investigated.
This validates that structural features of the considered geometrical setup are successfully incorporated via our first-principles functional learning.

Test particle sum rules serve as a further valuable thermodynamic consistency check.
We thereby find results for the excess chemical potential from different routes to agree numerically, which hence verifies the sum rules and the reliability of both the neural metadensity functional predictions and the employed functional calculus.
Capturing correctly the test particle thermodynamics is an important prerequisite for describing solvation free energies as well as the associated pertinent phenomena of drying and induced effective interactions \cite{WildingCriticalSurfacePhase2025}.

As we have demonstrated in this work, reduced geometries already provide substantial access to the rich phenomenology of inhomogeneous fluids, arguably surpassing the extent of insight that might be expected a priori.
Despite the evident advantages of exploiting symmetry from the outset, it would also be interesting in future work to consider extensions that allow one to work with full three-dimensional resolution.
Having access to efficient functionals in reduced (spherical and planar) geometry may thereby serve as reference in the construction of more sophisticated training protocols and neural network architectures.
In particular, the present study highlights the importance of efficient training data acquisition and handling; we recall that the spherical neural metadensity functional relied largely on radial distribution functions obtained in computationally inexpensive bulk fluid simulations, and that the metadensity functional concept allows well-controlled \cite{Dijkman2025LearningNeuralFreeEnergy,Ram2025LearnedFreeenergyFunctionals,Sammuller2024NeuralDensityFunctionals} incorporation in the training via Percus' test particle route.
Dealing efficiently with training data becomes particularly decisive if the spatial resolution is increased.
Our findings point to the relevance of exploiting particularly manageable simulation setups in lieu of more straightforward but arguably brute-force techniques, as would be the case e.g.\ when sampling volumetric data naively in simulations of three-dimensional fluids in general inhomogeneous landscapes.

Altering the neural network architecture is another promising measure in order to accomodate more generic geometries.
In particular, equivariant neural networks may mitigate technical difficulties that we expect to arise when dealing with volumetric data in a naive way.
These types of neural networks incorporate prescribed symmetry properies directly in their architecture, such that rotational symmetry, as is pertinent in the present work, could be encoded via $SO(3)$ equivariance.
In this regard, it should be emphasized that translation invariance may already be achieved e.g.\ with convolutional neural networks, as was recently leveraged in the context of classical density functional theory for two-dimensional hard disk systems \cite{Glitsch2025NeuralDensityFunctional}, and that recent applications to three-dimensional fluid models consider cubic point group equivariance \cite{Weimar2026CubicEquivariantNeural,Cheng2026EquivariantLearningTransferable}.

Further applications of the metadensity functional concept may consider different types of fluids and soft matter, in particular those featuring attractive or long-ranged interparticle interactions.
Combining previous insights on neural metadensity functionals with recent advances in the treatment of charged and molecular fluids \cite{Bui2025FirstprinciplesApproachElectromechanics,Bui2025LearningClassicalDensity,Bui2026UnifiedMachineLearning,Yang2025HighDimensionalOperator} may thereby alleviate the difficulty of capturing the oftentimes subtle and complex phenomenology that arises in such systems.
Addressing the connection of metadensity functional theory to the recent and closely related entropy density functional theory \cite{Schmidt2026EntropyDensityFunctionalTheory,Schmidt2026EntropyDensityFunctionalUniversality} constitutes a further valuable goal towards the systematic incorporation of pairwise interparticle interactions in a density functional framework.
Basing the machine learning on entropy density functional theory may provide a possible alternative to the present metadensity functional learning, which is to be investigated in future work.

\section*{Data Availability}

Code, data, and models that support the findings of this work are openly available \cite{GitHub}.
Calculations involving fundamental measure theory were carried out using the Julia package \verb|ClassicalDFT.jl| \cite{ClassicalDFT.jl}.

\begin{acknowledgments}
  Some of the calculations were performed using the festus-cluster of the Bayreuth Centre for High Performance Computing, funded by the DFG (Deutsche Forschungsgemeinschaft) under project no.~523317330.
  This work is supported by the DFG (Deutsche Forschungsgemeinschaft) under project no.~551294732.
\end{acknowledgments}

\appendix

\section{Pair direct correlation functions in the bulk fluid}
\label{appendix:c2spherical}

\begin{figure}[tb]
  \includegraphics[width=\columnwidth]{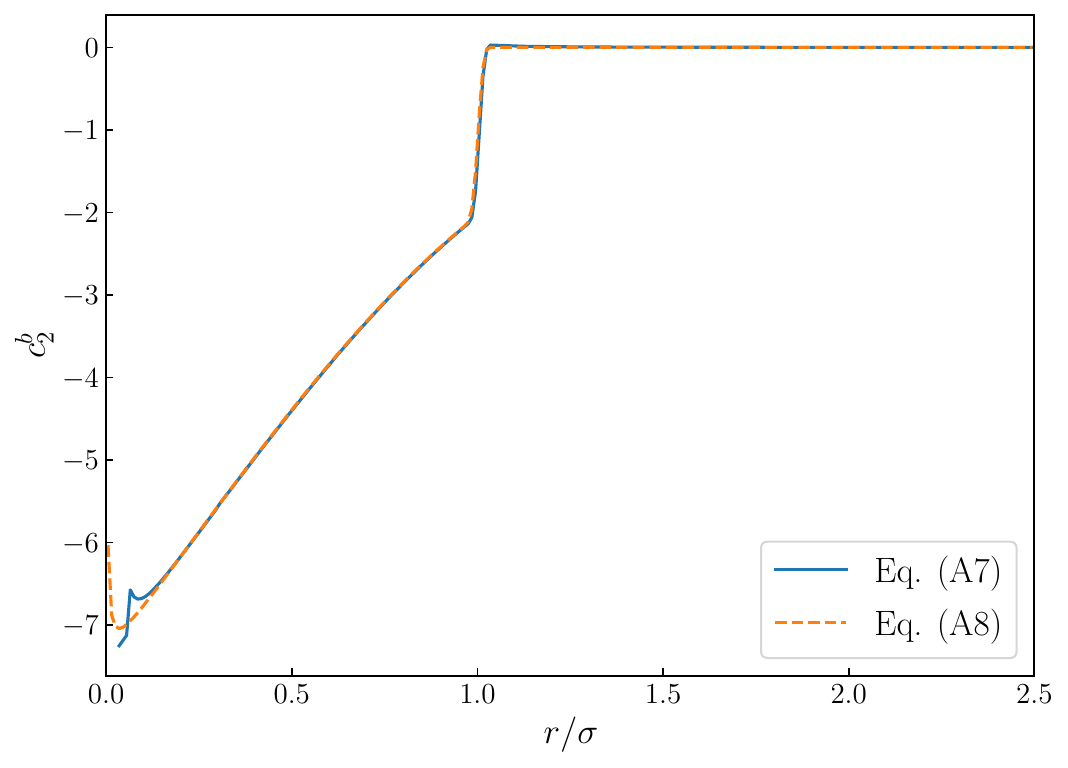}
  \caption{
    Verification of Eqs.~\eqref{EQc2br+r'} and \eqref{EQc2br-r'} using fundamental measure theory in Rosenfeld's formulation \cite{Rosenfeld1989FreeEnergyModel,Roth2010FundamentalMeasureTheory} for the underlying density functional.
    Numerical artifacts can be attributed to the required finite differencing and, particularly near the origin, to division by $r$.
  }
\label{FIGc2bFMT}
\end{figure}

To demonstrate the formal equivalence of Eqs.~\eqref{EQsphericalc2FourierIntegral} and~\eqref{EQsphericalc2RealSpaceIntegral}, we rewrite the latter via inserting therein the Fourier representation of $\tilde c_2^b(q)$ and re-ordering the sequence of the two integrals.
This yields
\begin{equation}
  \label{EQsphericalc2viaKernel}
  \sm{c}_2(r,r';\rho_b) = \int_0^\infty dr'' K(r,r',r'') c_2^b(r'') ,
\end{equation}
where
\begin{align}
  \begin{split}
    K(r,r',r'') &= \frac{1}{2\pi^2} \int_0^\infty dq q^2 \frac{\sin(q r)}{qr} \frac{\sin(q r')}{qr'}\\
                &\qquad\qquad\qquad \times 4 \pi r'' \frac{\sin(q r'')}{q}
  \end{split}\\
  &= \frac{2 r''}{\pi rr'}\int_0^\infty dq \frac{\sin(qr)}{q} \sin(qr')\sin(qr'')\\
  &= \begin{cases}
    \frac{r''}{2 r r'} & \text{if } |r-r'|< r'' < r+r'\\
    0 & \rm otherwise.
    \end{cases}
  \label{EQsphericalKernelSimplified}
\end{align}
Inserting Eq.~\eqref{EQsphericalKernelSimplified} into Eq.~\eqref{EQsphericalc2viaKernel} and simplifying leads to the relationship \eqref{EQsphericalc2AsPositionIntegral} between the spherical functional form of the two-body direct correlation function and its standard radial bulk version.
Comparing Eq.~\eqref{EQsphericalc2AsPositionIntegral} to the real space expression~\eqref{EQsphericalc2RealSpaceIntegral} yields the connecting variable substitution as $r''=\sqrt{r^2+r'^2-2urr'}$, which implies the differential relationship $du=-r''dr''/(rr')$.
The integral boundaries in Eq.~\eqref{EQsphericalc2RealSpaceIntegral} are $u=-1$ and $+1$ and these are mapped, respectively, in Eq.~\eqref{EQsphericalc2AsPositionIntegral} to $r''=r+r'$ and $|r-r'|$.
Exchanging the upper and lower limits of integration creates a minus sign that cancels the minus sign in the substitution of the differentials, thus proving the equivalence of Eqs.~\eqref{EQsphericalc2FourierIntegral} and~\eqref{EQsphericalc2RealSpaceIntegral}.

The simplified real space form \eqref{EQsphericalc2AsPositionIntegral} is further useful for recovering the bulk direct correlation function $c_2^b(r)$ as follows.
We calculate the first derivative of Eq.~\eqref{EQsphericalc2AsPositionIntegral} with respect to $r$ or $r'$, which yields, respectively,
\begin{align}
  \label{EQsphericalLinearEquation1}
  \begin{split}
    &rr' \frac{\partial \sm{c}_2(r,r'; \rho_b)}{\partial r} + r' \sm{c}_2(r,r';\rho_b) =\\
    &\quad\quad  \frac{r+r'}{2} c_2^b(r+r') - \frac{r-r'}{2} c_2^b(|r-r'|),
  \end{split}\\
  \label{EQsphericalLinearEquation2}
  \begin{split}
  &rr' \frac{\partial \sm{c}_2(r,r';\rho_b)}{\partial r'} + r \sm{c}_2(r,r';\rho_b) =\\
  &\quad\quad  \frac{r+r'}{2} c_2^b(r+r') + \frac{r-r'}{2} c_2^b(|r-r'|).
  \end{split}
\end{align}

The derivative relationships \eqref{EQsphericalLinearEquation1} and \eqref{EQsphericalLinearEquation2} constitute a set of two linear equations for $c_2^b(r+r')$ and $c_2^b(|r-r'|)$.
The solution has the following appealing form:
\begin{align}
  \label{EQc2br+r'}
  \begin{split}
    c_2^b(r+r') &= \sm{c}_2(r,r';\rho_b)\\
                &\quad+ \frac{rr'}{r+r'}\Big(\frac{\partial \sm{c}_2(r,r';\rho_b)}{\partial r} + \frac{\partial \sm{c}_2(r,r';\rho_b)}{\partial r'}\Big),
  \end{split}\\
  \label{EQc2br-r'}
  \begin{split}
    c_2^b(|r-r'|) &= \sm{c}_2(r,r';\rho_b)\\
                  &\quad- \frac{rr'}{r-r'}\Big(\frac{\partial \sm{c}_2(r,r';\rho_b)}{\partial r} - \frac{\partial \sm{c}_2(r,r';\rho_b)}{\partial r'}\Big).
  \end{split}
\end{align}
A practical caveat of the above expression is having to build the numerical derivatives with respect to the radial distance variables $r$ and $r'$ as well as the division by the sum or difference of $r$ and $r'$.

We verify in Fig.~\ref{FIGc2bFMT} results for $c_2^b(r)$ obtained from employing Eqs.~\eqref{EQc2br+r'} and \eqref{EQc2br-r'} using Rosenfeld's fundamental measure density functional theory \cite{Rosenfeld1989FreeEnergyModel}.
The corresponding one-body direct correlation functional $c_1(r; [\rho])$, applicable solely for hard spheres and hence without metadensity functional dependence, is thereby implemented according to the analytically reduced expressions in spherical geometry \cite{Roth2010FundamentalMeasureTheory,ClassicalDFT.jl}.
Following Eq.~\eqref{EQsphericalc2FunctionalDerivative}, automatic differentiation at spatially constant density profile $\rho(r) = \rho_b$ determines $\sph{c}_2(r, r';\rho_b)$, which in turn yields $\sm{c}_2(r, r';\rho_b)$ via Eq.~\eqref{EQsphericalc2bar} as required on the right hand sides of Eqs.~\eqref{EQc2br+r'} and \eqref{EQc2br-r'}.
The derivatives with respect to $r$ and $r'$ are evaluated numerically using finite differences.

\section{Inhomogeneous two-body correlation functions}
\label{appendix:two-body}

\begin{figure*}[tb]
  \includegraphics[width=\textwidth]{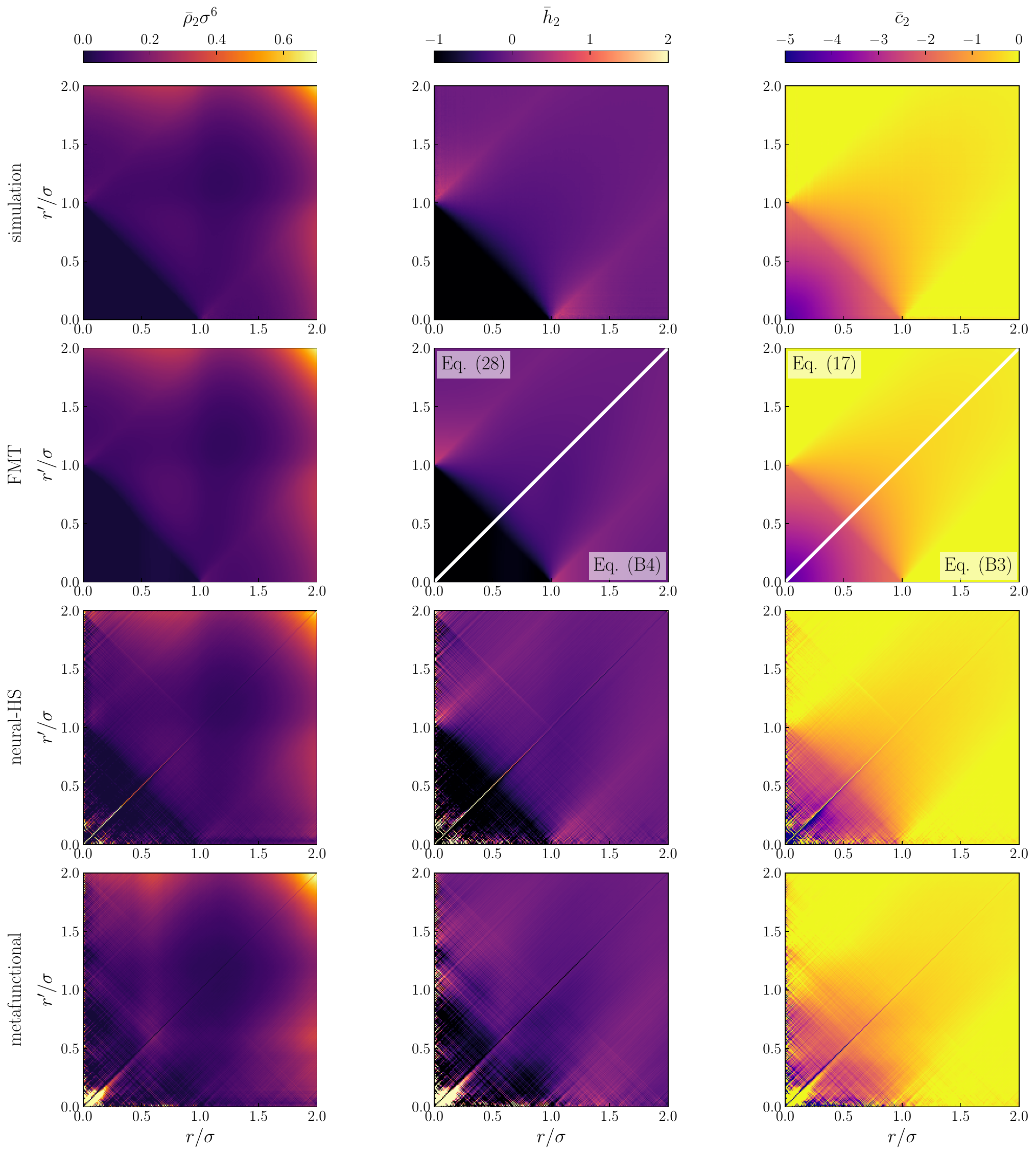}
  \caption{
    Spherically averaged two-body direct correlation functions for hard spheres confined in a spherical hard cavity of radius $2\sigma$.
    This corresponds to the situation shown in Figs.~\ref{FIGc2cavity} and \ref{FIGringc2cavity}; we investigate here the same routes (rows) and reproduce results for $\sm{c}_2(r, r')$ (third column).
    The spherically averaged two-body direct correlation functional $\sm{c}_2(r, r')$ is thereby converted via the spherical Ornstein-Zernike equation~\eqref{EQOZspherical} to $\sm{h}_2(r, r')$ (middle column) and via Eq.~\eqref{EQbarrho2} to $\sm{\rho}_2(r, r')$ (first column).
    For results obtained via fundamental measure theory, we additionally depict $\sm{c}_2(r, r')$ and $\sm{h}_2(r, r')$ as obtained from direct evaluation of Eqs.~\eqref{EQsphericalc2AsFunctionalDerivative} and \eqref{EQsphericalh2AsFunctionalDerivative} (below the diagonal) and observe agreement to the former Ornstein-Zernike results (above the diagonal) due to symmetry with respect to exchange of $r$ and $r'$.
    As a consistency check, we compare for the case of fundamental measure theory the result from Ornstein-Zernike inversion as well as explicit evaluation of functional derivative via numerical finite differencing according to Eqs.~\eqref{EQsphericalc2AsFunctionalDerivative} and \eqref{EQsphericalh2AsFunctionalDerivative}.
  }
\label{FIGrho2h2c2cavity}
\end{figure*}

In the following, the accessibility of further (direct and total) two-body correlation functions in spherically inhomogenous fluids is demonstrated.
We depict in Fig.~\ref{FIGrho2h2c2cavity} results for spherically averaged two-body quantities for the case of hard sphere confinement in a hard spherical cavity of radius $2\sigma$; see Figs.~\ref{FIGc2cavity} and \ref{FIGringc2cavity} for results of $\sph{c}_2(r, r')$ and $\sm{c}_2(r, r')$.
The spherically averaged Ornstein-Zernike equation \eqref{EQOZspherical} is used to compute the spherically averaged total correlation function $\sm{h}_2(r, r')$, from which the two-body density $\sm{\rho}_2(r, r')$ follows via Eq.~\eqref{EQbarrho2}.
This demonstrates the accessibility of both direct and total correlation functions on the basis of automatic differentiation of a (neural or analytical) density functional $c_1(r; [\rho])$.
For comparison, we consider $c_1(r; [\rho, \beta\phi])$ as represented by the neural metadensity functional as well as a neural functional without metadensity dependence trained purely for hard spheres.
Results are verified against fundamental measure theory, where $\sm{c}_2(r, r')$ follows from automatic differentiation of an analytical implementation of $c_1(r; [\rho])$, as well as against simulation results.

Equivalently to the Ornstein-Zernike relations laid out in the main text, one can express both the direct and total correlation function in terms of the following functional derivatives:
\begin{align}
  \label{EQc2AsFunctionalDerivative}
  c_2(\rv, \rv') &= \beta \frac{\delta V_\rmext(\rv)}{\delta \rho(\rv')} + \frac{\delta(\rv - \rv')}{\rho(\rv)},\\
  \label{EQh2AsFunctionalDerivative}
  h_2(\rv, \rv') &= \frac{-1}{\beta \rho(\rv) \rho(\rv')} \frac{\delta \rho(\rv)}{\delta V_\rmext(\rv')} - \frac{\delta(\rv - \rv')}{\rho(\rv)}.
\end{align}
The corresponding spherically averaged versions assume the form
\begin{align}
  \label{EQsphericalc2AsFunctionalDerivative}
  4 \pi r'^2 \sm{c}_2(r, r') &= \beta \frac{\delta V_\rmext(r)}{\delta \rho(r')} + \frac{\delta(r - r')}{\rho(r)}\\
  \label{EQsphericalh2AsFunctionalDerivative}
  4 \pi r'^2 \sm{h}_2(r, r') &= \frac{-1}{\beta \rho(r) \rho(r')} \frac{\delta \rho(r)}{\delta V_\rmext(r')} - \frac{\delta(r - r')}{\rho(r)}
\end{align}
and serve as valuable consistency checks in Fig.~\ref{FIGrho2h2c2cavity}, whereby the functional derivatives have been evaluated via finite differences.

\section{Functional line integration for test particle thermodynamics}
\label{appendix:testparticle}

As a consistency check for the test particle sum rules, we also consider an alternative functional integration method by parameterizing the functional line integral in density function space as $\rho_a(r) = \rho_b + a \Delta\rho(r)$, where the variable $0\leq a \leq 1$ and $\Delta \rho(r)=\rho_b g(r) - \rho_b$ is the difference of the test particle density profile $\rho_b g(r)$ and the bulk density $\rho_b$.
Using the standard splitting of the grand potential density functional, Eq.~\eqref{EQPercusMuExc} can be written as:
\begin{equation}
  \label{EQPercusThermodynamicAsFunctionalIntegral}
  \begin{split}
    c_1^b &= \Delta N +  \int d\rv \Delta\rho(r) \int_0^1 da  \Delta c_1(r;[\rho_a,\beta\phi])\\
                              &\quad -\int d\rv \rho_b g(r) [\beta\phi(r)  + \ln g(r)],
  \end{split}
\end{equation}
where the spatially integrated density difference is $\Delta N = \int d\rv \Delta\rho(r)$ and the difference of the one-body direct correlation functional and its bulk value is $\Delta c_1(r;[\rho,\beta\phi])= c_1(r;[\rho,\beta\phi]) - c_1(r;\rho_b,[\beta\phi])$.
Note that $c_1(r;\rho_b,[\beta\phi])=c_1^b$ and that $\Delta N = S(0)-1$, with the structure factor $S(k)$ evaluated at wavevector $k=0$.
Due to the spherical symmetry the position integrals are only radial, $\int d\rv \,\cdot\, = 4\pi \int_0^\infty dr r^2\,\cdot$.

Using that the general Euler-Lagrange equation attains the form $\Delta c_1(r;[\rho_b g,\beta\phi]) = \beta\phi(r) + \ln g(r)$ for test particle states, one can re-write Eq.~\eqref{EQPercusThermodynamicAsFunctionalIntegral} in the following `direct' way:
\begin{equation}
  \label{EQPercusThermodynamicAsFunctionalIntegralDirectVersion}
  \begin{split}
    c_1^b &= \Delta N -\int d\rv \rho_b g(r) \Delta c_1(r;[\rho_b g,\beta\phi])\\
                              &\quad +\int d\rv \Delta\rho(r) \int_0^1 da  \Delta c_1(r;[\rho_a,\beta\phi]).
  \end{split}
\end{equation}
For completeness, we re-collect the definitions of the objects that occur in Eq.~\eqref{EQPercusThermodynamicAsFunctionalIntegralDirectVersion}:
\begin{align}
  \Delta N &= \int d\rv \Delta\rho(r),\\
  \Delta \rho(r) &= \rho_b g(r) - \rho_b,
\end{align}
and furthermore:
\begin{align}
 \Delta c_1(r;[\rho,\beta\phi]) &= c_1(r;[\rho,\beta\phi]) - c_1(r;\rho_b,[\beta\phi]),\\
 \rho_a(r) &= \rho_b + a \Delta\rho(r).
\end{align}
Results obtained from this route are also shown in Fig.~\ref{FIGtpsumrule} and they coincide numerically with results of the former two routes.
This agreement hence validates the robust nature of thermodynamical predictions based on evaluation and functional line integration of the spherically trained neural metadensity functional.

\clearpage

\bibliography{spherical-neural.bib}

\end{document}